\documentclass[aps,prx,amsmath,amssymb,superscriptaddress,twocolumn]{revtex4}
\usepackage[english]{babel}
\usepackage{graphicx,bm,amssymb,amsmath}
\usepackage{color,wrapfig,braket}
\usepackage{hyperref,placeins,ulem}
\usepackage[caption=false]{subfig}
\usepackage{subfig}
\usepackage{ulem}

\usepackage{dsfont}
\usepackage{epsfig}
\usepackage{verbatim}
\usepackage{yhmath}
\usepackage{bbm}

\def\scE{\mathcal{E}}
\def\scM{\mathcal{M}}
\def\scO{\mathcal{O}}
\def\scS{\mathcal{S}}
\def\scX{\mathcal{X}}
\def\scZ{\mathcal{Z}}
\def\id{\mathds{1}}
\def\dd{\mathrm{d}}
\def\U{\mathrm{U}}

\def\Tr{\mathop{\mathrm{Tr}}}

\newcommand{\avg}{\mathop{\mathbb{E}}}

\newcommand{\eqnref}[1]{Eq.\,\eqref{#1}}
\newcommand{\figref}[1]{Fig.\,\ref{#1}}

\newcommand{\refcite}[1]{Ref.\,\onlinecite{#1}}

\newcommand{\dia}[3]{\raisebox{#3pt}{\includegraphics[height=#2pt]{dia_#1}}}

\makeatletter
\newcommand\xleftrightarrow[2][]{%
  \ext@arrow 9999{\longleftrightarrowfill@}{#1}{#2}}
\newcommand\longleftrightarrowfill@{%
  \arrowfill@\leftarrow\relbar\rightarrow}
\makeatother

\begin{document}

\title{Measurement-induced criticality in random quantum circuits }

\author{Chao-Ming Jian}
\affiliation{Station Q, Microsoft Quantum, Santa Barbara, California 93106-6105, USA}

\affiliation{Kavli Institute of Theoretical Physics, University of California, Santa Barbara, California 93106, USA}

\author{Yi-Zhuang You}
\affiliation{Department of Physics, University of California, San Diego, CA 92093, USA} 

\author{Romain Vasseur}
\affiliation{Department of Physics, University of Massachusetts, Amherst, MA 01003, USA}

\author{Andreas W. W. Ludwig}
\affiliation{Department of Physics, University of California, Santa Barbara, CA 93106, USA}

\date{\today}

\begin{abstract}

We investigate the critical behavior of the entanglement transition induced by projective measurements in (Haar) random 
unitary quantum circuits. Using a replica approach, we map the calculation of the entanglement entropies in such circuits onto a two-dimensional statistical mechanics model. In this language, the area- to volume-law entanglement transition can be interpreted as an ordering transition in the statistical mechanics model. We derive the general scaling properties of the entanglement entropies and mutual information near the transition using conformal invariance.  We analyze in detail the limit of infinite on-site Hilbert space dimension in which the statistical mechanics model maps onto percolation. In particular,
we compute the exact value of the universal coefficient of the logarithm of subsystem size in the $n$th R\'enyi entropies for $n \geq 1$ in this limit
using relatively recent results for conformal field theory describing the critical theory of 
2D percolation,
and we discuss how to access the generic transition at finite on-site Hilbert space dimension
from this limit, which is in a universality class different from 2D percolation.
We also comment on the relation to the entanglement transition in Random Tensor Networks, studied previously in~\refcite{Vasseur2018ETFHRTN}.

\end{abstract}

\maketitle

\section{Introduction}
Quantum entanglement plays a crucial role in modern condensed matter physics, both in equilibrium and non-equilibrium settings. Under unitary evolution, the entanglement of generic isolated many-body quantum systems tends to increase to a volume-law scaling of the entanglement entropies of subsystems~\cite{Calabrese_2005,PhysRevLett.111.127205,PhysRevLett.112.011601,Kaufman794,PhysRevB.95.094302,PhysRevX.7.031016,2018arXiv180300089J,PhysRevX.9.021033,2018arXiv181208657P}, as required by the  eigenstate thermalization hypothesis~\cite{PhysRevA.43.2046,PhysRevE.50.888} (ETH). It is then natural to ask whether different dynamical phases with different entanglement scaling can exist, and about the nature of the {\it entanglement transitions} separating these entanglement phases. An example of such an entanglement transition is provided by the many-body localization (MBL) transition~\cite{Luitz,PhysRevLett.113.107204,VHA,PVP,PhysRevX.5.041047,PhysRevX.7.021013,Schreiber842,DVP,THMdR_meanfield,zhang_many-body_2016,goremykina_analytically_2019,PhysRevB.99.094205}, which occurs in the presence of random or quasiperiodic potentials, and separates the volume-law thermal dynamical phase from an area-law (non-thermal) MBL phase~\cite{BAA,PhysRevB.75.155111,PalHuse,BauerNayak,PhysRevLett.111.127201,PhysRevB.90.174202,2014arXiv1404.0686N,RevModPhys.91.021001}.

A completely different way to obtain an entanglement transition
of a different kind
between area- and volume-law states was introduced in~\refcite{Vasseur2018ETFHRTN}. There the transition was induced by tuning the bond dimension of a state obtained at the boundary of a two-dimensional random tensor network. This entanglement transition can be described by an effective two-dimensional statistical mechanics model.

Shortly after, another type of entanglement transition was proposed using projective measurements: if a many-body quantum system is subjected to enough local measurements, such measurements can collapse the many-body wavefunction into an area-law entangled state, while with a low density of measurements, volume-law entanglement can survive. Such measurement-induced transitions in random unitary circuits~\cite{PhysRevX.7.031016,PhysRevX.8.021014,PhysRevX.8.021013,PhysRevX.8.041019,PhysRevLett.121.060601,PhysRevX.8.031058,PhysRevB.99.174205,PhysRevX.8.031057,2019arXiv190607736F} subjected to random local measurements were first introduced in Refs.~\cite{Skinner2018,LiChenFisher2018,Chan2019}, and were studied numerically both for Haar and Clifford random gates. 
Despite the growing interest in this transition~\cite{LiChenFisher2019,2019arXiv190305124C,2019arXiv190305452S, 2019arXiv190505195G}, it remains poorly understood, with the majority of results stemming from numerical observations. 
There have been only two exceptions:
(i) A fine-tuned transition between area and volume law entangled phases was 
shown in~\refcite{Vasseur2018ETFHRTN} to be in the
universality class of 
critical 2D percolation described by an exactly solvable conformal field theory (CFT). This provided an existence proof for such a transition.
Relaxing the fine-tuning induces a crossover to a transition in an analytically
so-far not tractable universality class.
(ii) Subsequently, the behavior of the zeroth R\'enyi entropy $S_0$ 
in the problem of the projective measurement-induced transition was mapped 
in Ref.~\onlinecite{Skinner2018} onto an 
exactly solvable ``geometric'' optimization problem for `minimal cuts' in 2D percolation.
Since in the same reference critical behavior of the zeroth R\'enyi entropy $S_0$
was observed at a parameter value (probability
of measurement) different from
the one where all $n$th R\'enyi entropies $S_n$ with $n\geq 1$ became critical,
the significance of the `minimal cut' results for $S_0$ for the measurement-induced
entanglement
transition remains to be better understood.

In this paper, we provide a theory of the projective measurement-induced entanglement transition (with Haar random unitary gates)
by mapping the calculation of entanglement entropies onto a statistical mechanics model. Our approach relies on a replica trick which allows us to deal with the intrinsic non-linearities of projective measurements. The area- to volume-law entanglement transition then corresponds to an ordering transition in the statistical mechanics model. This naturally explains the emergence of conformal invariance at the transition, and leads to universal scaling forms for the entanglement entropy and mutual information. In the limit of infinite on-site Hilbert space dimension $d=\infty$, we find that the entanglement transition is in the percolation universality class,
and we compute the exact value of the universal coefficient of the logarithm of subsystem size in 
all  $n$th R\'enyi entropies for $n\geq 1$
from the exactly known CFT, obtaining the value $=1/6$ for the entanglement of half of the system and open boundary conditions.
This is in contrast to the value of the universal coefficient
of the logarithm of subsystem size of the zeroth R\'enyi entropy computed in the same setting, as mentioned above, in Ref.~\onlinecite{Skinner2018} using the `minimal cut'
method, which was found in that work to be equal to $\approx 0.27$.
The fact that these two universal
numbers differ (by about a factor of two)
appears to indicate that, while in the limit of infinite on-site Hilbert space
dimension,  the  $n$th R\'enyi entropies for $n\geq 1$  and the zeroth
R\'enyi entropy $S_0$ happen to become critical at the {\it same} parameter 
value (probability of measurement), they describe rather different 
and unrelated properties of the 
system. (This is in line with the observation, mentioned above, that
these two quantities
become critical at different
parameter values in the generic case of {\it finite} on-site Hilbert space dimension.)
The limit of infinite on-site Hilbert space dimension
also allows us to identify the generic transition for finite on-site Hilbert space 
dimension as
that generated by a crossover from
the percolation conformal field theory by 
a single (Renormalization Group) relevant perturbation.

The remainder of this paper is organized as follows: in section~\ref{secRandomCircuits}, we introduce the model of random unitary circuits with random projective measurements, and explain how to compute the entanglement entropy using a replica approach.  In section~\ref{secStatmech}, we map the calculation of the entanglement entropy onto a statistical mechanics model, and 
discuss
the large $d$ limit. Section~\ref{secScaling} describes the consequences of conformal invariance for
scaling of various quantities for any $d$, while section~\ref{secPercolation} addresses the $d=\infty$ limit in detail. Finally, Sec.~\ref{secGeneric} deals with the nature of the transition at finite $d$ and the close relation to the entanglement transition~\cite{Vasseur2018ETFHRTN} in random tensor networks~\cite{Hayden2016,Qi:2017qf, Vasseur2018ETFHRTN}; and Sec.~\ref{secDiscussion} contains concluding remarks. 

\section{Random Quantum Circuits}
\label{secRandomCircuits}

\begin{figure}[t!]
\includegraphics[width=0.6\columnwidth]{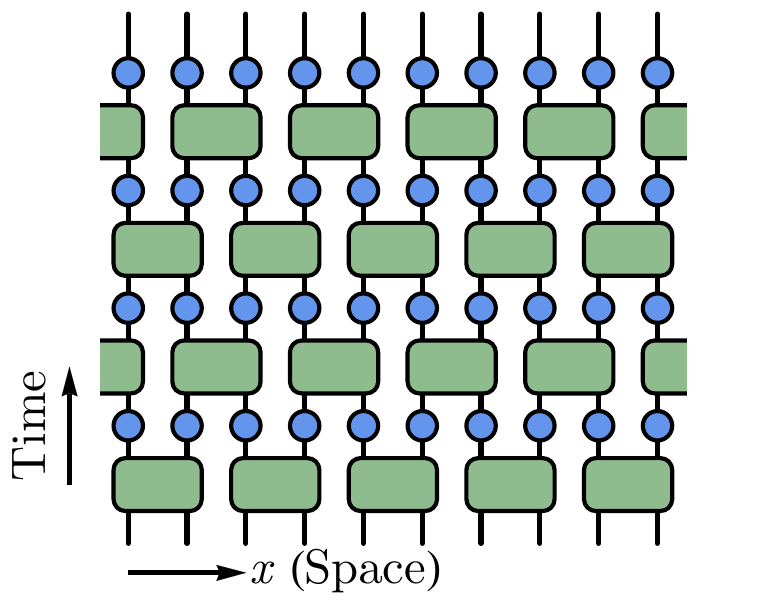}
\caption{Random unitary dynamics of a 1D qudit chain. The  blue circles represent one-site generalized measurements, while the green blocks represent Haar-random two-site unitary gates that act on pairs of neighbouring sites in the 1D qudit chain.  }
\label{Fig:RandomCircuit}
\end{figure}

We study the discrete-time dynamics of a 1D 
`qudit'  chain. That is, each site
of this 1D qudit chain has a local Hilbert space of dimension $d$. The discrete-time dynamics we focus on is generated by the quantum circuit with a ``brick-wall'' configuration shown Fig.~\ref{Fig:RandomCircuit} that consists of random unitary operators and generalized measurements. In Fig. \ref{Fig:RandomCircuit}, the 1D qudit chain is along the $x$ direction while the vertical direction represents time (or discrete time steps). Each green block represents an {\it independently Haar-random} two-site unitary gate that acts on a pair of neighbouring sites in the 1D qudit
chain.

Each of the blue blocks represents a one-site generalized measurement. Such generalized measurements can be most conveniently described using the language of quantum channels \cite{NielsenChuang,LiChenFisher2019}, which we review in the following. In general, a quantum channel is a completely positive trace-preserving map, which can be described by a set $\mathcal{M} = \{ M_\alpha \}$ of Kraus operators $M_\alpha $ (with $\alpha=1,2,...$). The Kraus operators are normalized according to a generalized normalization condition $\sum_{M_\alpha \in \mathcal{M}} w (M_\alpha) M_\alpha^\dag M_\alpha = \mathds{1}$ with $w (M_\alpha)$ a non-negative real number for each Kraus operator $M_\alpha \in \mathcal{M}$, which is the weight
of realizing $M_\alpha$ in the quantum channel. 
The left hand side of this normalization condition can be viewed as the weighted sum of $M_\alpha^\dag M_\alpha$'s with non-negative weights $w (M_\alpha)$. In the following, we will denote this weighted sum as $\avg_{M_\alpha \in \mathcal{M}}$. For example, we can rewrite the generalized normalization condition as $\avg_{M_\alpha \in \mathcal{M}} M_\alpha^\dag M_\alpha = \sum_{M_\alpha \in \mathcal{M}} w (M_\alpha) M_\alpha^\dag M_\alpha = \mathds{1}$. Given the set $\mathcal{M}$ and the weights, the quantum channel is defined as the map from any density matrix $\rho$ to $\avg_{M_\alpha \in \mathcal{M}}  M_\alpha \rho M_\alpha^\dag$. In fact, in the standard definition of the Kraus operators and their normalization (see \cite{NielsenChuang} for example), the weights $w(M_\alpha)$ are all taken to be 1. Here, we have made a generalization to non-unity weights and to the corresponding weighted sum for the convenience of later discussion. Given the set $\mathcal{M}$ (and the weights of the Kraus operators), the quantum channel can also be understood as a ``probabilistic evolution". If one starts with a pure quantum state $|\psi \rangle $, for every Kraus operator $M_\alpha\in \mathcal{M} $, the quantum channel evolves $|\psi \rangle $ to $ \frac{M_\alpha |\psi \rangle}{\| M_\alpha |\psi \rangle \|}$ with a probability of $w(M_\alpha) \| M_\alpha |\psi \rangle \|^2 = w(M_\alpha) \langle \psi | M_\alpha^\dag M_\alpha |
\psi \rangle$. Note that this probability is normalized due to the generalized normalization condition of the Kraus operators. Since we only consider one-site generalized measurements in the quantum circuit shown in \figref{Fig:RandomCircuit}, we then restrict the Kraus operators in $\mathcal{M}$ to be localized on the site where the corresponding blue block is acting on.

The quantum channel description of generalized measurements can easily recover the standard projective measurement. For example, the one-site projective measurement with respect to a (orthonormal)
set of
basis vectors $|i\rangle$ (with $i=1,2,...,d$)  of the $d$ dimensional local Hilbert space on a given site can be described by the quantum channel with the set of Kraus operators $\mathcal{M}_1 = \{P_1, P_2, ..., P_d \}$ and the weights $w(P_i) = 1$ for $i=1,2,...,d$. Here, $P_i = |i\rangle \langle i |$ is the 
projection operator on the
$i$th basis vector. The quantum channel with the set $\mathcal{M}_1$ evolves (or collapses) a pure state $|\psi\rangle$ to $\frac{P_i |\psi\rangle}{\|P_i |\psi\rangle \|}$ with a probability of $w(P_i)\|P_i |\psi\rangle \|^2 = \|P_i |\psi\rangle \|^2 $, which is as expected for the standard projective measurement.

Using the language of quantum channels, one can study more generalized forms of measurements. Ref.~\onlinecite{Skinner2018} and Ref.~\onlinecite{Chan2019} studied quantum circuits with $d=2$ in similar configurations as Fig. \ref{Fig:RandomCircuit} in which a quantum state, when it encounters a blue block in the quantum circuit, undergoes a standard one-site projective measurement with a classical probability $p$ and stays intact with a classical probability $1-p$.
In this scenario, the associated quantum channel is described by the set of Kraus operators $\mathcal{M}_p = \{ \mathds{1}, P_1, P_2,... , P_d \}$ equipped with the weights $w(\mathds{1})=1-p$ and $w(P_i)=p$ for $i=1,2,...,d$. In the following sections, we also study this type of generalized measurement (or quantum channel) given by the set of Kraus operators
$\mathcal{M}_p$ and the corresponding weights given above. We would also like to introduce a closely related generalized measurement given by the set of Kraus operators
$\mathcal{M}_p' = \{\mathds{1} \} \cup  \{\sqrt{d} P_U | U\in \U(d) \}  $ with $P_U\equiv U^\dag P_1 U$, which is an 
(uncountable)
infinite set. The
subset $ \{\sqrt{d} P_U | U\in \U(d) \}  $ of $\mathcal{M}_p' $ is continuously parameterized by a (one-site) unitary matrix $U\in \U(d) $. The weight on the operator $\mathds{1} \in \mathcal{M}_p'$ is still $1-p$, which has the same physical interpretation as the weight of the operator $\mathds{1}$ in $\mathcal{M}_p$. The weight on the infinite subset $ \{\sqrt{d} P_U | U\in \U(d) \}  $ is naturally given by the Haar measure: $w(\sqrt{d} P_U) = p\, \dd U $ where $\dd U$ represents the Haar measure on $\U(d)$,  normalized such that $\int_{U\in \U(d)} \dd U \ \mathds{1} =\mathds{1}$. The weighted sum of Kraus operators in $\mathcal{M}_p'$ is defined accordingly. For example, the generalized normalization condition of Kraus operator is given by $\avg_{M \in \mathcal{M}_p'}M^\dag M = (1-p) \mathds{1}^\dag \mathds{1} + p \int_{U\in \U(d)} \dd U (\sqrt{d} P_U)^\dag (\sqrt{d} P_U) = \mathds{1}$. The relation between the generalized measurements defined by $\mathcal{M}_p$ and $\mathcal{M}_p'$ will be studied in the following section.

Before we focus on a specific choice of generalized measurements, let us rephrase the construction of the quantum circuit of interest to us in this paper using the quantum-channel language we introduced above. In a random quantum circuit of the configuration shown in Fig. \ref{Fig:RandomCircuit}, each green block is an independently Haar-random two-site unitary gate that acts on a pair of neighboring sites in the 1D qudit chain. Each of the blue blocks is independently and randomly drawn from the ensemble given by the set of Kraus operators
$\mathcal{M}$ (and the associated weights) that is associated with the
generalized measurement one wants to study. For each realization of the green and blue blocks, we can built a random quantum circuit, denoted as $C$, following \figref{Fig:RandomCircuit}. Such a quantum circuit evolves an initial pure state $|\psi \rangle$ of the 1D qudit chain to the pure state $\frac{C|\psi \rangle}{\|C|\psi \rangle\| } $. The probability for this evolution to occur is the product of three factors:
(i) the norm
$\|C|\psi \rangle\|^2 = \langle\psi| C^\dag C | \psi \rangle = \Tr(C | \psi \rangle \langle\psi| C^\dag )$,
(ii) the weight for each Kraus operator in each blue block, and
(iii) the Haar-measure probability for
realizing each random two-site unitary gates in each green block. In \refcite{Wiseman1996}, the evolution from $|\psi \rangle$ to $\frac{C|\psi \rangle}{\|C|\psi \rangle\| } $ is also referred as to a {\it quantum trajectory}. Different realizations of the random quantum circuit $C$ lead to different quantum trajectories.

We are interested in the average quantum dynamics induced by this random quantum circuit. We denote the average over all realizations of the random quantum circuit as $\avg_C \cdots$. The precise meaning of $\avg_C$ is the following. First, for each two-site random unitary gate (green block), $\avg_C$ contains an independent integration over the Haar measure (of the $\mathrm{U}(d^2)$ group). This integration will be denoted as $\avg_U$ in the following. For each generalized measurement (blue block), $\avg_C$ includes the weighted sum over the set of Kraus operators,
$\avg_\mathcal{M}$, as explained before.

One example 
of an averaged quantity under the quantum dynamics induced by this random quantum 
circuit is the averaged expectation value $\bar{\mathcal{O}}$ of an observable $\mathcal{O}$ 
in the state
obtained from evolving the initial state $|\psi\rangle$ by the random quantum circuit:
\begin{align}
\bar{\mathcal{O}} & = \avg_C \left( \frac{ \langle\psi | C^\dag \mathcal{O} C|\psi \rangle}{\|C|\psi \rangle\|^2 } \times   \Tr(C | \psi \rangle \langle\psi| C^\dag )  \right)   
 \nonumber \\
 & = \avg_C  \langle\psi | C^\dag \mathcal{O} C|\psi \rangle ,
 \label{Eq:Operator_AveC}
\end{align} 
where $\frac{ \langle\psi | C^\dag \mathcal{O} C|\psi \rangle}{\|C|\psi \rangle\|^2 }$ is the quantum mechanical
expectation value of the observable $\mathcal{O}$ 
in the state $\frac{C|\psi \rangle}{\|C|\psi \rangle\| } $, and the factor $ \Tr(C | \psi \rangle \langle\psi| C^\dag ) $ is, as explained, one of the factors
in the probability of the corresponding quantum trajectory. We see that Eq. \ref{Eq:Operator_AveC} naturally agrees with the evolution of an observable $\mathcal{O}$ under a quantum channel (constructed from Fig. \ref{Fig:RandomCircuit} by viewing each of the green and blue blocks as a quantum channel). The more interesting quantity we want to study is the (averaged) dynamics of subsystem entanglements under the random quantum circuit. Consider a subsystem $A$ and its 
complement $\bar{A}$ of the 1D qudit chain. The $n$th R\'enyi entropy $S_{n,A}[|\psi\rangle]$ on the subsystem $A$ of a pure state $|\psi\rangle$ follows the standard definition:
\begin{align}\label{eq:def S}
S_{n,A}[|\psi\rangle] = \frac{1}{1-n} \log\Tr_A \left( \rho_A^n \right),
\end{align}
where $\rho_A = \Tr_{\bar{A}} | \psi \rangle \langle \psi|$ is the reduced density matrix on the subsystem $A$. Here, $\Tr_A$ (or  $\Tr_{\bar{A}}$) represents the partial trace over the degrees of freedom in the subsystem $A$ (or $\bar{A}$). 
Alternatively, $S_{n,A}[|\psi\rangle]$ in \eqnref{eq:def S} can be expressed in terms of the expectation value of a permutation operator
$\scS_{n,A}$ acting on the the $n$-fold replicated state as
\begin{equation}
S_{n,A}[|\psi\rangle]=\frac{1}{1-n}\log\Tr\left((\ket{\psi}\bra{\psi})^{\otimes n}\scS_{n,A}\right),
\end{equation}
where $\scS_{n,A}$ depends on the choice of entanglement region $A$ and is defined as
\begin{equation}\label{eq:boundary def}
\scS_{n,A}=\prod_{x}\scX_{g_x},\quad
g_x=\left\{\begin{array}{ll}(12\cdots n), & x\in A,\\
{\rm identity} = e,
& x\in \bar{A}.\end{array}\right.
\end{equation}
$g_x$ labels the permutation on site $x$, and $\scX_{g_x}=\sum_{[i]}\ket{i_{g_x(1)}i_{g_x(2)}\cdots i_{g_x(n)}}\bra{i_1i_2\cdots i_n}$ is its representation on the replicated on-site Hilbert space,
i.e. on its $n$-fold tensor product~\footnote{An othonormal basis of which
is denoted by $|i_1, i_2, ..., i_n\rangle=$
$|i_1\rangle \otimes |i_2\rangle \otimes ... \otimes |i_n\rangle$.}. 
Here, as indicated in the equation above, $g_x$ is the cyclic (identity) permutation when $x$ is in the region $A$ (when $x$ is in the region $\bar{A}$). We are interested in the averaged $n$th R\'enyi entropy $\bar{S}_{n,A}$ of the final state after the random quantum circuit evolution:
\begin{align}
\bar{S}_{n,A} =  \avg_C S_{n,A}\left[\frac{C|\psi \rangle}{\|C | \psi \rangle\| } \right]   \times   \Tr(C | \psi \rangle \langle\psi| C^\dag ) , 
\label{Eq:Ave_SnA1}
\end{align}
which can be rewritten as 
\begin{widetext}
\begin{equation}
\begin{split}
\bar{S}_{n,A} &=  \frac{1}{1-n}\avg_C \left(\log \frac{\Tr\left((C\ket{\psi}\bra{\psi}C^{\dagger})^{\otimes n}\scS_{n,A}\right)}{\Tr(C | \psi \rangle \langle\psi| C^\dag)^{\otimes n}}\right)\Tr(C | \psi \rangle \langle\psi| C^\dag),\\
&=\lim_{m \rightarrow 0} \frac{1}{m(1-n)} \avg_C\left(\left(\Tr(C\ket{\psi}\bra{\psi}C^{\dagger})^{\otimes n}\scS_{n,A}\right)^m-\left(\Tr(C | \psi \rangle \langle\psi| C^\dag)^{\otimes n}\right)^m\right)\Tr(C | \psi \rangle \langle\psi| C^\dag),\\
&=\lim_{m \rightarrow 0} \frac{1}{m(1-n)} \avg_C \Tr\left((C\ket{\psi}\bra{\psi}C^{\dagger})^{\otimes nm+1}(\scS_{n,A}^{\otimes m}-\id)\right),
\label{Eq:Ave_SnA2}
\end{split}
\end{equation}
\end{widetext}
where we have introduced a second replica index $m$ to resolve the ensemble average of the logarithm using
$\log x=\lim_{m\to 0}(x^m-1)/m$ and $\Tr(X^{\otimes m})= (\Tr X)^m$. This
replica trick was introduced in~\refcite{Vasseur2018ETFHRTN} in the context of random tensor networks, and in
~\refcite{PhysRevB.99.174205} for random unitary circuits. In this double replica scheme, the total number of replica is $Q=nm+1$, and the replica limit $m\to 0$ corresponds to $Q\to 1$. As can be seen from \eqnref{Eq:Ave_SnA2}, the additional replica apart from $nm$ originated from the  probability $\Tr(C | \psi \rangle \langle\psi| C^\dag)$ of obtaining a measurement outcome. As a side comment, if we re-weight this probability by a power $q$, i.e.~replacing $\Tr(C | \psi \rangle \langle\psi| C^\dag )\to \Tr(C | \psi \rangle \langle\psi| C^\dag )^q$, we could also realize other replica limits $Q=mn+q\to q$ as $m\to0$.

To evaluate \eqnref{Eq:Ave_SnA2}, we will need to calculate the ensemble average $\avg_C C^{\otimes Q}\otimes C^{\dag\otimes Q} $ of
the tensor product of
$Q$ copies of the random quantum circuit $C$ and $Q$ copies of its conjugate $C^\dag$.
In the next section, we will show that the calculation of the average $\avg_C  C^{\otimes Q} \otimes C^{\dag\otimes Q} $ can be mapped onto a statistical mechanics model in 2+0 dimensions. By imposing different boundary conditions corresponding to fixing permutations at the boundary [following Eq. (\ref{eq:boundary def})] to $\scS_{n,A}^{\otimes m}$ or $\id$, the statistical mechanics model results in different partition functions
\begin{equation}\label{eq:Z def}
\begin{split}
\scZ_A&=\avg_C \Tr\left((C\ket{\psi}\bra{\psi}C^{\dagger})^{\otimes Q}\scS_{n,A}^{\otimes m}\right),\\
\scZ_\emptyset&=\avg_C \Tr(C\ket{\psi}\bra{\psi}C^{\dagger})^{\otimes Q},
\end{split}
\end{equation}
from which the averaged $n$th R\'enyi entropy $\bar{S}_{n,A}$ can be obtained in the replica limit via
\begin{equation}
\bar{S}_{n,A}=\frac{n}{1-n}\lim_{Q\to 1}\frac{\scZ_{A}-\scZ_\emptyset}{Q-1}.
\end{equation}
Using the fact that $\scZ_A= \scZ_\emptyset = 1$ in the replica limit $m \to 0$ ($Q \to 1$), this can be rewritten in a more intuitive form as the free energy cost of the domain-wall associated with changing the boundary condition in the entanglement region:
\begin{equation}
\label{eqFenergycost}
\bar{S}_{n,A}=\lim_{m \to 0}\frac{F_{A}-F_\emptyset}{m(n-1)}=\lim_{m \to 0}\frac{\log(\scZ_A/\scZ_\emptyset)}{m(1-n)},
\end{equation}
with $F_A = - \log \scZ_{A}$ and $F_\emptyset = - \log \scZ_{\emptyset}$.
Also, from the  statistical-mechanics-model perspective, we will see that the choice of the initial state $|\psi \rangle$ is not essential when the depth, namely the number of discrete time steps, of the random quantum circuit becomes large.

\section{Statistical Mechanics Model}
\label{secStatmech}

Let us derive the statistical mechanics model for generic replica number $Q$ first, before taking the replica limit $Q\to 1$.
The evaluation of the expectation value
$\avg_C C^{\otimes Q}\otimes C^{\dag\otimes Q}$ boils down to the ensemble average of the unitary gates and the generalized measurements
in the circuit. Let $U$ be a two-site Haar-random unitary gate. The average of the tensor product of $Q$ identical copies of $U$ and $U^\dag$ under the Haar measure is given by (using standard graphical notations~\cite{ORUS2014117}):
\begin{equation} \label{eqHaaraverage}
\avg_U \dia{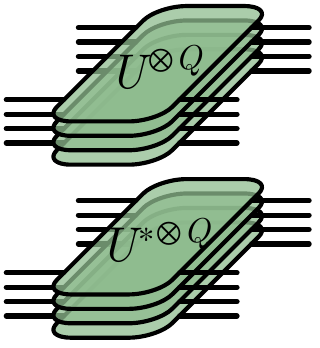}{56}{-26}=\sum_{g_1,g_2\in S_Q}\mathsf{Wg}_{d^2}(g_1^{-1}g_2)\dia{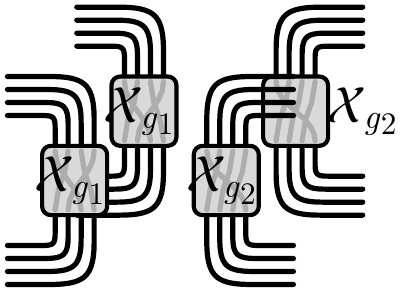}{47}{-20},
\end{equation}
where $\mathsf{Wg}_{D}(g)$ denotes the Weingarten function of the permutation $g$, 
\begin{equation}\label{eq:Wg def}
\mathsf{Wg}_D(g)=\frac{1}{Q!}\sum_{\lambda\vdash Q}\frac{\chi_\lambda(e)\chi_\lambda(g)}{\prod_{(i,j)\in Y(\lambda)}(D-i+j)},
\end{equation}
where the sum is taken over all integer partitions $\lambda$ of $Q$ [denoted in the above equation by the notation $\lambda\vdash Q$, such that $\lambda = (\lambda_1, \lambda_2, ...)$ with $\lambda_1\geq \lambda_2\geq\cdots$, $\lambda_i\in\mathbb{N}$ and $\sum_i\lambda_i=Q$],
and the product is taken over all cells $(i,j)$ in the Young diagram $Y(\lambda)$ of the shape $\lambda$. Here $e$ denotes the identity group element, and $\chi_{\lambda}(g)$ is the irreducible character of the symmetric group $S_Q$ indexed by the partition $\lambda$.

As we average over all two-site unitary gates in the circuit, the partition function will break up into a product of independent contributions from the generalized measurements. Each generalized measurement is associated with the following partition function weight
\begin{equation}
W_M(g_1,g_2)=\dia{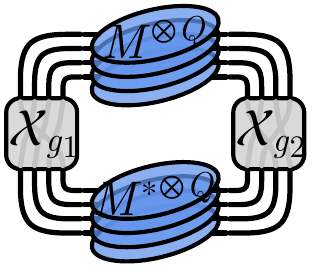}{43}{-19}=\Tr\scX_{g_1}M^{\otimes Q}\scX_{g_2}M^{\dagger\otimes Q},
\end{equation}
where $M$ is an element of a set of Kraus operators.
For $M=\id$, we have $W_\id(g_1,g_2)=\Tr\scX_{g_1}\scX_{g_2}=d^{|g_1^{-1}g_2|}$, where $|g|$ 
denotes~\footnote{The quantity $|g|$ was denoted in~\refcite{Vasseur2018ETFHRTN}
by the symbol $C(g)$. The quantity ($|e|$-$|g|$), where $e$ denotes the identity permutation, equals the minimal number of transpositions necessary to represent
the permutation $g$ as a product of transpositions.} the number of cycles in the permutation $g$ (including cycles of length $1$). We consider two scenarios of the generalized measurement, described by the previously discussed sets of Kraus operators $\scM_p$ and $\scM'_p$, respectively. They lead to seemingly different ensemble averages: 
\begin{equation}\label{eq:W over Mp}
\avg_{M\in\scM_p}W_M(g_1,g_2)=(1-p) d^{|g_1^{-1}g_2|}+p d,
\end{equation}
\begin{equation}\label{eq:W over Mpp}
\avg_{M\in\scM'_p}W_M(g_1,g_2)
=(1-p) d^{|g_1^{-1}g_2|}+p d^{Q},
\end{equation}
However, in the replica limit $Q\to 1$, the two scenarios converge to the same partition function weight, although we emphasize that these factors can matter if one considers the limit $d \to \infty$ before the replica limit $Q\to 1$. We will take \eqnref{eq:W over Mpp} for generic $Q$, and define the following weight function
\begin{equation}
\label{LabelEqGenericBoltzmannWeight}
W_p(g)=(1-p) d^{|g|}+p d^{Q},
\end{equation}
which will be useful in constructing the Boltzmann weight of the following statistical mechanics model.

\begin{figure}[t!]
\begin{center}
\includegraphics[width=0.95\columnwidth]{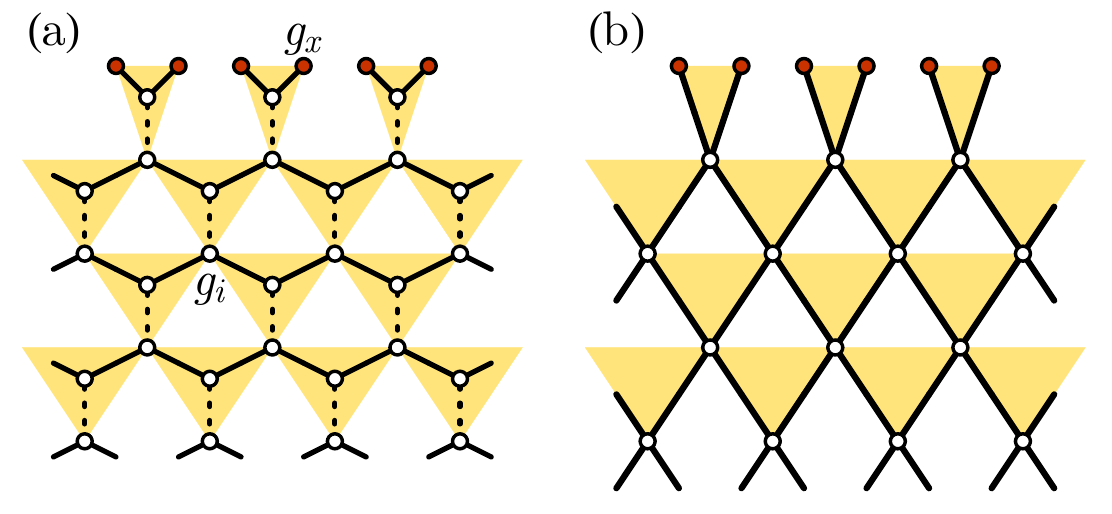}
\caption{(a) Geometry of the statistical mechanics model of $S_Q$ spins. The red sites corresponds to the boundary spins to be pinned by the boundary condition. (b) In the $d=\infty$ limit, the model reduces to a Potts model on a square lattice.}
\label{fig:lattice}
\end{center}
\end{figure}

Put together, the partition function $\scZ_A$ in \eqnref{eq:Z def} can be formulated as a statistical mechanics model on an anisotropic honeycomb lattice as depicted in \figref{fig:lattice}(a) where a permutation group element
$g_i\in S_Q$, a  `spin', is defined on each site,
\begin{equation}\label{eq:ZA}
\scZ_A=\sum_{\{g_i\in S_Q\}}\prod_{\langle ij\rangle\in\scE_s}W_p(g_i^{-1}g_j)\prod_{\langle ij\rangle\in\scE_d}\mathsf{Wg}_{d^2}(g_i^{-1}g_j),
\end{equation}
and
where
$\scE_s$ ($\scE_d$) denotes the set of solid (dotted) links
on the lattice. In connection with the original network geometry of the random circuit in \figref{Fig:RandomCircuit}, the vertical (dotted) links on the honeycomb lattice represent the Weingarten functions which originated from averaging the two-site unitary gates, and the zigzag (solid) links keep track of the contributions from the generalized measurements. 
In the following, we will refer to the model described by
\eqnref{eq:ZA} as the ``$S_Q$ model''.
A similar statistical mechanics model was first derived in \refcite{PhysRevB.99.174205} for Haar
random unitary circuits without projective
measurements (i.e.~$p=0$). Here we generalized the model to the case with projective measurement.

We note a crucial symmetry property of
the statistical mechanics model that
will become important in our discussion below, arising from the following
symmetry of the local weights
$W_p(g_i^{-1}g_j)$
and
$\mathsf{Wg}_{d^2}(g_i^{-1}g_j)$ which enter the partition function
in Eq. (\ref{eq:ZA}): They are invariant under global right- and left-multiplication of all group elements,
\begin{eqnarray}
\label{LabelEqGlobalSymmetry}
g_i \to h_L g_i h_R^{-1},  g_j \to h_L g_j h_R^{-1},
\   {\rm where} \ h_L, h_R \in S_Q.
\end{eqnarray}
This invariance follows from the fact that 
the Weingarten function in Eq. (\ref{eq:Wg def})
as well as the `cycle-counting function'
which appears in
Eq. (\ref{LabelEqGenericBoltzmannWeight})
and
assigns to each permutation $g$
the number of its cycles $|g|$,
are both `class functions' (i.e. depend only the conjugacy class
of the permutation group element).

The $S_Q$ `spins' on the boundary, which are permutation group elements
$g_x\in S_Q$ for boundary sites $x$,
are pinned by the boundary condition, 
which is specified 
by the entanglement region $A$ as follows:
\begin{equation}\label{eq:SQ boundary}
g_x=\left\{\begin{array}{ll}g_{\rm SWAP} \equiv (12\cdots n)^{\otimes m}, & x\in A,\\ 
{\rm identity} = e,
& x\in\bar{A}.\end{array}\right.
\end{equation}
This equation follows from \eqnref{eq:boundary def} by taking into account the $m$ replica which arise, as discussed above, in addition to the R\'enyi replica from rewriting averages of the logarithm. By tuning the probability of measurement
$p$, we can change the couplings on the solid links and drive,
as we will see, an entanglement transition. As will be shown below, this measurement-induced transition can be naturally interpreted as a simple symmetry-breaking transition of the statistical mechanics model.

In the limit 
where
the on-site Hilbert space dimension $d=\infty$ is infinite, the $S_Q$ model
turns out to reduce
to a Potts model with $Q!$ colors. To see this, we evaluate the partition function
weight $J_p(g_i,g_j;g_k)$ associated with
each down triangle (in yellow) in \figref{fig:lattice}(a),
\begin{equation}\label{eq:J def}
\begin{split}
&J_p(g_i,g_j;g_k)=\sum_{g_l\in S_Q}\dia{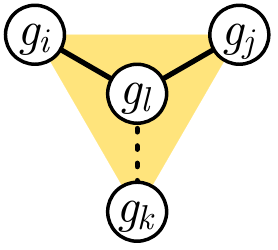}{40}{-20}\\
&=\sum_{g_l\in S_Q}W_p(g_i^{-1}g_l)W_p(g_j^{-1}g_l)\mathsf{Wg}_{d^2}(g_l^{-1}g_k).
\end{split}
\end{equation}
The partition function in \eqnref{eq:ZA} can be equivalently written in terms of the triangle weight $J_p$ as
\begin{equation}\label{eq:ZA inf d}
\scZ_A=\sum_{\{g_i\in S_Q\}}\prod_{\langle ijk\rangle\in\triangledown}J_p(g_i,g_j;g_k),
\end{equation}
subject to the boundary condition that $g_i$ should match $g_x$ as specified in \eqnref{eq:SQ boundary} on the boundary. Note that positivity of this weight is not guaranteed in general. For example
$J_{p=0}((123),(132);e)=-2(d^2-1)/(d^6+d^4-4d^2-4)<0$ (for any realistic on-site Hilbert space dimension
$d\geq 2$). This makes the statistical mechanics non-unitary, which is not an issue for our approach, as the field theory describing the entanglement transition is necessarily non-unitary 
in any case
because of the replica limit (see below). In the $d\to\infty$ limit, we obtain
\begin{equation}\label{eq:inf d}
\begin{split}
W_p(g)&=d^{Q}((1-p)\delta_g+p),\\
\mathsf{Wg}_{d^2}(g)&=d^{-2Q}\delta_g,
\end{split}
\end{equation}
where $\delta_g$ is the delta function that gives 1 if and only if $g=e$ is the identity element in the permutation group $S_Q$, and gives 0 otherwise. A detailed derivation of \eqnref{eq:inf d} can be found in Appendix~\ref{app:1/d}. Substituting~\eqnref{eq:inf d} into \eqnref{eq:J def}, the triangle weight reduces to
\begin{equation}\label{eq:J inf d}
J_p(g_i,g_j;g_k)=((1-p)\delta_{g_i^{-1}g_k}+p)((1-p)\delta_{g_j^{-1}g_k}+p),
\end{equation}
which further factorizes into partition function weights defined separately on the bonds $\langle ik\rangle$ and $\langle jk\rangle$. The partition function weight across the bond $\langle ik\rangle$
equals
$1$ if $g_i=g_k$ and $p$ if $g_i\neq g_k$, and  an
analogous weight is assigned to the bond $\langle jk\rangle$. If we treat each on-site group element $g_i\in S_Q$ as a state (color) in a spin model, the partition function weight in \eqnref{eq:J inf d} precisely matches that of a $Q!$-state Potts model on a square 
lattice, whose links are between sites $i$ and $k$, and between
sites $i$ and $j$ in each unit cell, as displayed in \figref{fig:lattice}(b).
By tuning the measurement rate $p$, the partition function $\scZ_A$ in \eqnref{eq:ZA inf d} undergoes a phase transition from the ordered phase (small $p$) to the disordered phase (large $p$) which we will analyze in detail below. 

Away from the $d=\infty$ limit, the weight $W_p(g)$ receives the following leading corrections $W_p(g)=d^{Q}((1-p)\delta_g+p+\frac{1-p}{d}\delta'_{g}+\scO(d^{-2}))$, where $\delta'_g=1$ if $g$ is a transposition such as e.g. $(12)$, and $\delta'_g=0$ otherwise. The  Weingarten function will not receive corrections at $1/d$ order. Using these results, 
the triangle weight can be evaluated to the $1/d$ order (see Appendix~\ref{app:1/d} for details), yielding
\begin{equation}\label{eq:J to 1/d}
\begin{split}
J_p(g_i,g_j;g_k)&=((1-p)\delta_{g_i^{-1}g_k}+p)((1-p)\delta_{g_j^{-1}g_k}+p)\\
&+\frac{1-p}{d}\big((1-p)\delta'_{g_i^{-1}g_j}(\delta_{g_i^{-1}g_k}+\delta_{g_j^{-1}g_k})\\
&\phantom{+\frac{1-p}{d}(}+p(\delta'_{g_i^{-1}g_k}+\delta'_{g_j^{-1}g_k})\big)+\scO(d^{-2}).
\end{split}
\end{equation}
Moreover,  up to this order,
we find that the weights of the $S_Q$ model factorize into
a product of weights associated with bonds of the 
same square lattice, depicted in \figref{fig:lattice}(b), 
that appears in the $d=\infty$ result~\eqref{eq:J inf d}. Denoting the local weight on the bond $\langle ik\rangle$ as $e^{- E(g_i, g_k)}$, the energy function reads (see Appendix~\ref{app:1/d})  
\begin{eqnarray} \label{eqEnergyLink}
&&
E(g_i,g_k) =
\\ \nonumber
&&
=
- \log \left[ p + (1-p) \left(\delta_{g_i^{-1}g_k}+ \frac{1}{d}\delta'_{g_i^{-1}g_k} \right) + \scO(d^{-2}) \right].
\end{eqnarray}
We see that, among all the domain walls, the $1/d$ corrections favor energetically transposition domain walls in our model, Eq. (\ref{eq:ZA inf d}).
--- this will turn out to have important consequences in the following. Crucially, these $1/d$ corrections break the artificially large $S_{Q!}$ symmetry of the weights~\eqnref{eq:J inf d}
of
the $d\to \infty$ limit to the $S_Q\times S_Q$ symmetry present for 
finite $d$. Consequences of this reduction of symmetry
by the $1/d$ corrections will be further analyzed below.

\section{Conformal invariance}

\label{secScaling}
Now that we have mapped the calculation of the entanglement entropies of the random circuit with projective measurements onto a (replica) statistical mechanics model, many qualitative features of the entanglement transition can be understood naturally. Our discussion follows closely~\refcite{Vasseur2018ETFHRTN} where a similar statistical mechanics model was found to describe entanglement transitions in random tensor networks. At small $p$, the Boltzmann weights give a ferromagnetic interaction favoring  group elements (`spins') on neighboring sites to be equal,
and we thus expect an ordered phase of the statistical mechanics model. 
In that phase, 
the free energy cost  $F_A-F_\emptyset$ in \eqnref{eqFenergycost} associated with ``twisting'' the entanglement region scales with the size $L_A$ of $A$ at long times (many layers in the circuit), corresponding to volume-law entanglement $ \bar{S}_{n,A} \sim L_A$ for sufficiently deep circuits (in the long-time limit $t \to \infty$). This is clearly the behavior expected without measurement, i.e. at $p=0$. As the measurement rate $p$ gets closer to 1, the effective temperature of the statistical mechanics model is increased, leading to a disordered phase. 
The domain wall condensate present in this phase
can freely absorb the domain wall at the boundaries of the entanglement interval, such that, for a distance exceeding the correlation length from the boundary, there is no additional free energy cost from the boundary domain. In this limit, the free-energy cost of the boundary domain will scale like the boundary of $A$, corresponding to area-law scaling of entanglement $ \bar{S}_{n,A} \sim \text{const}$. 

The entanglement transition separating these two phases therefore corresponds to an ordering transition in the statistical mechanics model.  In general, assuming that the transition is of second order, it should be described a by two-dimensional Conformal Field Theory (CFT) with central charge $c=0$ in the replica limit $Q \to 1$. (Recall that $c$ measures the way the free energy changes when a finite scale is introduced; since here the partition function $\scZ_{\emptyset}=1$ is trivial in the replica limit, we have $c=0$.) 
Such CFTs at central charge $c=0$ are non-unitary, and
are notoriously hard to tackle even in two dimensions. Below 
we will propose a way to approach this transition from the infinite on-site Hilbert space size $d=\infty$ limit.
Even without identifying the underlying CFT precisely, there are important consequences that can be deduced from conformal invariance alone.  

First of all, since the bulk properties of the transition only depend on $Q$, the location of the bulk transition point at
$p=p_c$ 
as well as all bulk critical exponents are the same for all R\'enyi entropies in the replica limit $Q \to 1$. (The R\'enyi entropies arise from observables located at the boundary of the
system.)
Our statistical model thus naturally explains why all R\'enyi entropies with $n \geq 1$ have a transition at the same value of $p_c$. (This was observed numerically in Ref. \cite{Skinner2018}.)

Obviously, conformal invariance implies a dynamical critical
exponent $z=1$, so the scaling with time and space should be the same at the entanglement transition. To analyze the scaling of the entanglement entropy at the critical point, we note that the ratio of partition functions  $\scZ_{A}/\scZ_{\emptyset}$ that appears in \eqnref{eqFenergycost}, corresponds in the CFT language to the two-point function of a boundary condition changing (BCC) 
{operator~\cite{cardy_conformal_1984,cardy_boundary_2006}
$\phi_{\rm BCC}$:
\begin{eqnarray}
\label{LabelEqDefBccOperator}
&&
\scZ_{A}/\scZ_{\emptyset} = \langle \phi_{\rm BCC}(L_A) \phi_{\rm BCC}(0) \rangle,
\end{eqnarray}
where the operators are inserted at  the boundary of the entanglement interval $A$. Near criticality, this two-point function scales as $\sim 1/L_A^{2 h (n,m)} f_{n,m} (L_A/\xi_{Q})$ with $\xi_Q \sim \left| p - p_c(Q) \right|^{-\nu(Q)}$ the correlation length of the statistical mechanics model and  $f_{n,m}$ are universal scaling functions that depend on $n$ and $m$ independently. Plugging this expression into the replica formula~\eqref{eqFenergycost}, we find the general scaling of the entanglement entropy
\begin{equation}
\bar{S}_{n,A}  =   \frac{2}{n-1} \left.\frac{\partial h}{\partial m} \right|_{m=0} \log L_A + f_n \left( \frac{L_A}{\xi}\right),
\label{eqEntanglementLog}
\end{equation}
with $\xi \sim \left| p - p_c\right|^{-\nu}$ the correlation length in the limit $Q\to 1$. In particular, conformal invariance predicts that $\bar{S}_{n,A}  \sim \log L_A$ at criticality $p=p_c$, with a universal prefactor that depends on the R\'enyi index $n$. 
Note that eq.~\eqref{eqEntanglementLog} holds up to additive non-universal constants  --- in order to isolate the universal contributions, one can also take the derivative of $\bar{S}_{n,A}$ with respect to $\log L_A$. This scaling form is in good agreement with the numerical observations of Refs.~\onlinecite{Skinner2018,LiChenFisher2018,LiChenFisher2019}.

The BCC operator $\phi_{\rm BCC}$ can also be used to derive the scaling of the mutual information of two regions $A=[x_1,x_2 ]$ and $B=[x_3,x_4 ]$: ${\cal I}^{n}_{A,B} = \bar{S}_{n,A}+ \bar{S}_{n,B} -  \bar{S}_{n,A \cup B}$, which maps naturally onto the 4-point function of $\phi_{\rm BCC}$. As a result of conformal invariance, we find that the mutual information at criticality should  depend only on the cross ratio~\cite{francesco2012conformal}
\begin{equation}
{\cal I}^{n}_{A,B} = g_n (\eta) \text{ with } \eta=\frac{x_{12} x_{34}}{x_{13} x_{24}} ,
\end{equation}
where
$x_{ij} = \frac{L}{\pi} \sin \frac{\pi}{L} |x_i - x_j|$ for a system of size $L$ with periodic boundary conditions. This scaling was checked numerically for Clifford unitary
circuits in~\refcite{LiChenFisher2019}.

\section{Percolation limit $d = \infty$ }
\label{secPercolation}
As we have shown above, the Boltzmann weights of the statistical model in the limit of infinite onsite Hilbert space dimension $d \to \infty$ take a very simple form~\eqnref{eq:J inf d}. This coincides with the high temperature expansion of a Potts model with $Q!$ states on the square lattice, as an expansion onto the so-called Fortuin-Kasteleyn clusters~\cite{FORTUIN1972536} where an edge is occupied with weight $1-p$, not occupied with weight $p$, and where each connected cluster (including single sites) carries a weight $Q!$ --- the number of Potts
states. In the replica limit $Q \to 1$, this maps onto a bond percolation problem where $1-p$ is the probability for a bond to be occupied. The partition function is trivial $\scZ_\emptyset = 1$, and the transition occurs for $p_c=1/2$. The correlation length diverges as $\xi \sim \left| p - p_c\right|^{-4/3}$ at the transition~\cite{PhysRevB.23.429}, and the central charge is $c=0$ as expected from general considerations. 

As just discussed, we have shown that the entanglement transition driven by projective measurements is in the universality class of 2D percolation 
($Q \to 1$ limit of a $Q!$-Potts model). We note that based on earlier
results on random unitary circuits without measurement at infinite on-site Hilbert space dimension~\cite{PhysRevX.7.031016,PhysRevB.99.174205}, and 
based on the description obtained in
Ref.~\onlinecite{Skinner2018} for the zeroth Renyi entropy
in terms of a minimal cut classical optimization problem of paths in 2D percolation, it was conjectured in Ref.~\onlinecite{Skinner2018} that if the
minimal cut classical optimization problem holds exactly in the projective measurement problem in the limit of infinite Hilbert space dimension $d$, then this optimization problem would also describe the $n$th Renyi entropies with
$n\geq 1$, and not only the zeroth Renyi entropy $S_0$, with the same result. While Ref.~\onlinecite{Skinner2018}  thus anticipated, based on these previous works, a connection of the projective measurement problem with percolation in the
$d \to \infty$ limit, there are universal quantities that go beyond this minimal cut picture and which can only be captured using a detailed analysis that relies on our replica trick formulation, as well
as on detailed properties of the CFT describing the percolation critical point.
In particular, we will show below that the $d = \infty$ limit requires a detailed knowledge of the CFT of 2D percolation, including very recent results~\cite{Dubail_2010}, rather than merely a geometric ``optimization problem'' as in~\refcite{Skinner2018} for $S_0$.

To illustrate this point, we 
now provide an exact calculation of
the universal prefactor of the logarithm in \eqnref{eqEntanglementLog} in the limit $d \to \infty$. To do so, we need to identify the proper BCC operator in the $Q!$-state Potts CFT. 
This is actually a subtle point: Naively, this would  appear to 
be the BCC operator which changes the boundary condition  that
is fixed to the identity permutation group element, to the boundary condition that is fixed to the 
``SWAP'' group element [2nd line of Eq.
(\ref{eq:SQ boundary})] of the $Q!$-state Potts model in the percolation limit 
$Q\to 1$. It is well known that this BCC operator has finite scaling dimension 
$=1/3$ in that limit~\cite{cardy_conformal_1984,CARDY1989581,Cardy_1992}. This would imply an infinite limit $m\to 0$  for
 all R\'enyi entropies in Eq. (\ref{eqFenergycost}) using a powerlaw in Eq. (\ref{LabelEqDefBccOperator}) with a finite decay exponent $=2\times 1/3$ in that limit.

This issue with this naive approach arises because
the limit $d \to \infty$ was taken implicitly before the replica limit $Q\to 1$. To remedy this,
the key idea is to ``soften'' the Boltzmann weights of the statistical mechanics model
in the vicinity of the boundary in a small ``boundary layer'' and replace them
by those at  a finite value of $1/d$. The bulk Boltzmann weights remain at $1/d=0$,
i.e. they are those of the Potts model. Since the Boltzmann weights of the boundary layer
still favor ``ferromagnetic'' alignment of the $S_Q$-valued `spins', the presence of the 
boundary layer does not modify the bias for boundary `spins' to align to the SWAP  and the identity group elements, respectively, along segments $A$ and $\bar{A}$ of the boundary.
The effect of the boundary layer is that the `sharp' domain wall where
the group element along the boundary switches directly from identity to SWAP, 
splits~\footnote{The notion of splitting of domains walls characterized by permutation
group elements was previously discussed in the different context of entanglement
in many-body chaotic {\it unitary} Haar random circuits, without measurements, in~\refcite{PhysRevB.99.174205}.} into a sequence of
$m(n-1)$  consecutive ``elementary'' domain walls, each characterized by a single transposition 
having just one cycle of length two since domain walls
with a single transposition 
are energetically favored by the finite-$1/d$
correction in the
energy function \eqnref{eqEnergyLink}. Using eq.~\eqnref{eqEnergyLink}, it is straightforward to see that the energy cost
of an elementary domain wall is $\Delta E_{\rm elementary} = \log p^{-1} - \frac{1-p}{p} d^{-1} + \scO(d^{-2}) $, which is lower than the energy cost of the domain wall separating the identity permutation 
from SWAP, which has energy
$\Delta E_{\rm SWAP} = \log p^{-1}+ \scO(d^{-2})
$.
Note also that the total energy cost of an extended  segment of the boundary
separating uniform boundary conditions fixed to the identity on one side from
uniform boundary conditions fixed to SWAP on the other of this segment, which consists of a sequence of
$m(n-1)$ consecutive domain walls (whose group theory product must be equal
to the SWAP group element), is also less than the cost of a `sharp'
SWAP domain wall located on  a single boundary link, 
since $m(n-1) \Delta E_{\rm elementary} \ll \Delta E_{\rm SWAP} $ in the replica limit $m \to 0$.
Moreover, since 
the energy cost
$\Delta E_{\rm elementary}$ of a single transposition domain wall on a given boundary link is lower than that of a domain wall on the same link characterized by
any other non-identity permutation [using \eqnref{eqEnergyLink}], the sharp
SWAP domain wall localized at a single boundary link will split into
$m(n-1)$ elementary domains walls, each  localized
on one of the $m(n-1)$  boundary links.

While the so-defined $(n-1)m$ elementary domain walls
in the 2D Potts model can branch and touch each other, they are well-known and well-defined objects in the 2D Potts model, called `thin' domain walls, whose
properties have recently been studied in great detail~\cite{Dubail_2010}. In our
context the corresponding  `split' BCC operator inserts $\ell=m(n-1)$ `thin'
domain walls in the Potts theory.
Using the results of~\refcite{Dubail_2010}, we find that the relevant BCC operator is $\Phi_{\rm BCC}=\Phi_{2 \ell -1, 4 \ell -1}$, using standard CFT notations~\cite{francesco2012conformal}~\footnote{More precisely, we identify $\Phi_{\rm BCC}$ as the most relevant operator in the fusion of the ``fixed/fixed'' BCC operator $\Phi_{1,3}$ with $\Phi_{1+2(\ell-1),1+4(\ell-1)} $. The latter operator inserts $\ell -1$ domain walls (since $\Phi_{1,3}$ already insert one domain wall)  with otherwise free boundary conditions, as introduced in Ref.~\cite{Dubail_2010}.  }. Now, the $Q!$-state Potts model is described by a CFT~\cite{PhysRevB.23.429,PhysRevB.27.1674} with central charge $c=1-\frac{6}{x(x+1)}$ and $x=\frac{\pi}{\arccos \frac{\sqrt{Q!}}{2}} - 1$. The scaling dimension of the boundary operator  $\Phi_{r,s}$ is then $h_{r,s} = \frac{((x+1) r -s x)^2-1}{4 x (x+1)}$.
The critical exponent $h_{2 \ell -1, 4 \ell -1}$ vanishes as $\ell = m (n-1) \to 0$ in the replica limit $m\to 0$, as it should,
and yields 
$\lim_{m\to 0} h_{2 \ell -1, 4 \ell -1}/(n-1)m=1/6$.
In the replica limit, ~\eqnref{eqEntanglementLog} therefore yields
(for periodic spatial boundary conditions)}
\begin{equation}
\bar{S}_{n,A}  =   \frac{1}{3}\log L_A + \dots,
\label{eqEntanglementLogPercolation}
\end{equation}
for all $n$th R\'enyi entropies $n \geq 1$ at criticality in the limit $d \to \infty$. 
We remark that our replica statistical mechanics  model
only describes R\'enyi entropies with index $n \geq 1$, as quantities such as the domain wall free energy all change sign at $n=1$.
While, as already mentioned above, 
 Ref.~\onlinecite{Skinner2018}  anticipated, based on  previous works, 
 a connection of the projective measurement problem with percolation in the
$d \to \infty$ limit,
we emphasize that the universal prefactor in eq.~\eqref{eqEntanglementLogPercolation} goes beyond
the geometric `minimal cut' path optimization picture found in
Ref.~\onlinecite{Skinner2018}
to describe the zeroth R\'enyi entropy,
and  indicates a different behavior of the R\'enyi entropies $n \geq 1$. We also note  that the universal 
prefactor in~\eqref{eqEntanglementLogPercolation} is not purely a property of the 2D percolation CFT, as it depends on how this CFT is approached in the replica limit  $m \to 0$--- see~\eqref{eqEntanglementLog}.   

We end by commenting that the same expression for the universal coefficient of the logarithm of subsystem size can be obtained in the Random Tensor Network model of~\refcite{Vasseur2018ETFHRTN},  when fine-tuned to the percolation critical point.

\section{Generic entanglement transition at finite Hilbert space dimension $d$}

\label{secGeneric}

In closing, we briefly comment on the CFT describing of
the generic 
entanglement transition at finite Hilbert space dimension $d$. While the percolation limit $d \to \infty$ provides an in essence  completely
analytically tractable
picture of the projective measurement-induced entanglement transition, this limit is not generic. The Potts model which is obtained in the limit $d \to \infty$ has a symmetry $S_{Q!}$, which is much larger than the $S_Q \times S_Q$ symmetry (corresponding to left/right multiplication) of  the model~\eqref{eq:ZA} describing finite $d$.
The leading operator in the $S_{Q!}$ Potts model that breaks the symmetry down to $S_Q \times S_Q$ was identified in~\refcite{Vasseur2018ETFHRTN} as the so-called two-hull operator of the Potts model.  In the replica percolation limit $Q \to 1$,  this operator has scaling dimension $\Delta_{\rm 2-hull}=\frac{5}{4} < 2$ so it is relevant.
(In fact, it turns out
that this is the only Renormalization Group relevant operator
that can appear at the percolation fixed point, when the symmetry is broken
to $S_Q \times S_Q$.)
We therefore expect the finite $d$ entanglement transition to be described by a different CFT, obtained as the IR fixed point of percolation perturbed by a two-hull operator. More precisely, let us work with the Landau-Ginzburg formulation of the $Q!$-state Potts field theory in terms of the Potts order parameter field
$ \phi_a $ with $a=1,\dots,Q!$, and $\sum_a \phi_a =0$. The leading perturbation implementing the symmetry breaking $S_{Q!} \longrightarrow S_Q \times S_Q $ is given by~\cite{Vasseur2018ETFHRTN}
\begin{equation}
{\cal L} ={\cal L}_{\rm Potts} + \sum_{a,b \in S_{Q}} W(a^{-1} b) \phi_a \phi_b + \dots
\end{equation}
where $W$ is a class function of the permutation group
$S_Q$. Crucially the labels $a,b$ are now interpreted as elements of the group $S_Q$. The only allowed function $W(a^{-1} b)$ that would respect the $S_{Q!}$ 
symmetry is $W(a^{-1}b)=\delta_{a,b}$, but any
class function of $S_Q$ is enough to satisfy the  $S_Q \times S_Q$ symmetry. The fate of this perturbed Potts model in the IR is
currently
unknown. However, we note that this field theory has exactly the same form as the one obtained for entanglement transitions in bulk random tensor networks in~\refcite{Vasseur2018ETFHRTN}, as they both correspond to the symmetry breaking $S_{Q!} \longrightarrow S_Q \times S_Q $.  It is therefore tempting to conjecture that they correspond to the same bulk universality class, although we caution that the two transitions correspond to different replica limits; $Q\to 1$ and $Q \to 0$ respectively, for the projective measurement transition studied in this work and for random tensor networks, respectively. In both cases, $Q! \to 1$ in the replica limit, so both problems correspond to a percolation theory perturbed by a 2-hull operator. However, the different replica limits will likely yield different
(trivial) prefactors, {\it e.g.} in~\eqnref{eqEntanglementLog}.

\section{Discussion}
\label{secDiscussion}

We have derived an exact statistical mechanics model description of the entanglement transition that occurs in random unitary circuits with projective measurements~\cite{Skinner2018,LiChenFisher2018}. Our approach relies on a replica trick that allows us to average entanglement entropies over the realizations of the random circuits and measurements, and to deal with the intrinsic non-linearities of the projective measurement problem. Our work naturally explains the emergence of conformal invariance at the entanglement transition, and predicts specific scaling forms for the entanglement entropy and mutual information. In the limit of infinite Hilbert space dimension $d=\infty$, we find that the transition is in the percolation universality class, and we computed the exact value of the universal coefficient of the logarithm of subsystem size in the $n$th R\'enyi entropies for $n\geq 1$. This limit also provides a natural starting point to identify the generic entanglement transition at finite Hilbert space dimension $d$, as a perturbation of percolation by a 2-hull operator. Identifying the CFT in the generic case remains an challenging task for future work. 

{\it Note.}--- While we were finalizing
this manuscript, a related work appeared on the ArXiv~\cite{2019arXiv190804305B}. This work also derives a statistical mechanics model for the transition driven by weak measurements, but the conclusions regarding the $d \to \infty$ and finite $d$ limits appear to be different from ours.  

{\it Acknowledgments.}--- We thank M.P.A. Fisher, S. Gopalakrishnan, D. Huse, A. Nahum and A.C. Potter
for insightful discussions. We are grateful to  the KITP, which is supported by the National Science Foundation under Grant NSF PHY-1748958, and the KITP Program ``The Dynamics of Quantum Information'' where parts of this work were carried out. This work is supported in part by the National Science Foundation under Grant DMR-1309667 (AWWL), the Gordon and Betty
Moore Foundation's EPiQS Initiative through Grant No.
GBMF4304 (CMJ), the US Department of Energy, Office of Science, Basic Energy Sciences, under Early Career Award No. DE-SC0019168 (RV), and the Alfred P. Sloan Foundation through a Sloan Research Fellowship (RV). 

\bibliography{CT}

\begin{thebibliography}{62}
\expandafter\ifx\csname natexlab\endcsname\relax\def\natexlab#1{#1}\fi
\expandafter\ifx\csname bibnamefont\endcsname\relax
  \def\bibnamefont#1{#1}\fi
\expandafter\ifx\csname bibfnamefont\endcsname\relax
  \def\bibfnamefont#1{#1}\fi
\expandafter\ifx\csname citenamefont\endcsname\relax
  \def\citenamefont#1{#1}\fi
\expandafter\ifx\csname url\endcsname\relax
  \def\url#1{\texttt{#1}}\fi
\expandafter\ifx\csname urlprefix\endcsname\relax\def\urlprefix{URL }\fi
\providecommand{\bibinfo}[2]{#2}
\providecommand{\eprint}[2][]{\url{#2}}

\bibitem[{\citenamefont{{Vasseur} et~al.}(2018)\citenamefont{{Vasseur},
  {Potter}, {You}, and {Ludwig}}}]{Vasseur2018ETFHRTN}
\bibinfo{author}{\bibfnamefont{R.}~\bibnamefont{{Vasseur}}},
  \bibinfo{author}{\bibfnamefont{A.~C.} \bibnamefont{{Potter}}},
  \bibinfo{author}{\bibfnamefont{Y.-Z.} \bibnamefont{{You}}}, \bibnamefont{and}
  \bibinfo{author}{\bibfnamefont{A.~W.~W.} \bibnamefont{{Ludwig}}},
  \bibinfo{journal}{arXiv e-prints} \bibinfo{eid}{arXiv:1807.07082}
  (\bibinfo{year}{2018}), \eprint{1807.07082}.

\bibitem[{\citenamefont{Calabrese and Cardy}(2005)}]{Calabrese_2005}
\bibinfo{author}{\bibfnamefont{P.}~\bibnamefont{Calabrese}} \bibnamefont{and}
  \bibinfo{author}{\bibfnamefont{J.}~\bibnamefont{Cardy}},
  \bibinfo{journal}{Journal of Statistical Mechanics: Theory and Experiment}
  \textbf{\bibinfo{volume}{2005}}, \bibinfo{pages}{P04010}
  (\bibinfo{year}{2005}),
  \urlprefix\url{https://doi.org/10.1088%2F1742-5468%2F2005%2F04%2Fp04010}.

\bibitem[{\citenamefont{Kim and Huse}(2013)}]{PhysRevLett.111.127205}
\bibinfo{author}{\bibfnamefont{H.}~\bibnamefont{Kim}} \bibnamefont{and}
  \bibinfo{author}{\bibfnamefont{D.~A.} \bibnamefont{Huse}},
  \bibinfo{journal}{Phys. Rev. Lett.} \textbf{\bibinfo{volume}{111}},
  \bibinfo{pages}{127205} (\bibinfo{year}{2013}),
  \urlprefix\url{https://link.aps.org/doi/10.1103/PhysRevLett.111.127205}.

\bibitem[{\citenamefont{Liu and Suh}(2014)}]{PhysRevLett.112.011601}
\bibinfo{author}{\bibfnamefont{H.}~\bibnamefont{Liu}} \bibnamefont{and}
  \bibinfo{author}{\bibfnamefont{S.~J.} \bibnamefont{Suh}},
  \bibinfo{journal}{Phys. Rev. Lett.} \textbf{\bibinfo{volume}{112}},
  \bibinfo{pages}{011601} (\bibinfo{year}{2014}),
  \urlprefix\url{https://link.aps.org/doi/10.1103/PhysRevLett.112.011601}.

\bibitem[{\citenamefont{Kaufman et~al.}(2016)\citenamefont{Kaufman, Tai, Lukin,
  Rispoli, Schittko, Preiss, and Greiner}}]{Kaufman794}
\bibinfo{author}{\bibfnamefont{A.~M.} \bibnamefont{Kaufman}},
  \bibinfo{author}{\bibfnamefont{M.~E.} \bibnamefont{Tai}},
  \bibinfo{author}{\bibfnamefont{A.}~\bibnamefont{Lukin}},
  \bibinfo{author}{\bibfnamefont{M.}~\bibnamefont{Rispoli}},
  \bibinfo{author}{\bibfnamefont{R.}~\bibnamefont{Schittko}},
  \bibinfo{author}{\bibfnamefont{P.~M.} \bibnamefont{Preiss}},
  \bibnamefont{and} \bibinfo{author}{\bibfnamefont{M.}~\bibnamefont{Greiner}},
  \bibinfo{journal}{Science} \textbf{\bibinfo{volume}{353}},
  \bibinfo{pages}{794} (\bibinfo{year}{2016}), ISSN \bibinfo{issn}{0036-8075},
  \eprint{https://science.sciencemag.org/content/353/6301/794.full.pdf},
  \urlprefix\url{https://science.sciencemag.org/content/353/6301/794}.

\bibitem[{\citenamefont{Ho and Abanin}(2017)}]{PhysRevB.95.094302}
\bibinfo{author}{\bibfnamefont{W.~W.} \bibnamefont{Ho}} \bibnamefont{and}
  \bibinfo{author}{\bibfnamefont{D.~A.} \bibnamefont{Abanin}},
  \bibinfo{journal}{Phys. Rev. B} \textbf{\bibinfo{volume}{95}},
  \bibinfo{pages}{094302} (\bibinfo{year}{2017}),
  \urlprefix\url{https://link.aps.org/doi/10.1103/PhysRevB.95.094302}.

\bibitem[{\citenamefont{Nahum et~al.}(2017)\citenamefont{Nahum, Ruhman, Vijay,
  and Haah}}]{PhysRevX.7.031016}
\bibinfo{author}{\bibfnamefont{A.}~\bibnamefont{Nahum}},
  \bibinfo{author}{\bibfnamefont{J.}~\bibnamefont{Ruhman}},
  \bibinfo{author}{\bibfnamefont{S.}~\bibnamefont{Vijay}}, \bibnamefont{and}
  \bibinfo{author}{\bibfnamefont{J.}~\bibnamefont{Haah}},
  \bibinfo{journal}{Phys. Rev. X} \textbf{\bibinfo{volume}{7}},
  \bibinfo{pages}{031016} (\bibinfo{year}{2017}),
  \urlprefix\url{https://link.aps.org/doi/10.1103/PhysRevX.7.031016}.

\bibitem[{\citenamefont{{Jonay} et~al.}(2018)\citenamefont{{Jonay}, {Huse}, and
  {Nahum}}}]{2018arXiv180300089J}
\bibinfo{author}{\bibfnamefont{C.}~\bibnamefont{{Jonay}}},
  \bibinfo{author}{\bibfnamefont{D.~A.} \bibnamefont{{Huse}}},
  \bibnamefont{and} \bibinfo{author}{\bibfnamefont{A.}~\bibnamefont{{Nahum}}},
  \bibinfo{journal}{arXiv e-prints} \bibinfo{eid}{arXiv:1803.00089}
  (\bibinfo{year}{2018}), \eprint{1803.00089}.

\bibitem[{\citenamefont{Bertini et~al.}(2019)\citenamefont{Bertini, Kos, and
  Prosen}}]{PhysRevX.9.021033}
\bibinfo{author}{\bibfnamefont{B.}~\bibnamefont{Bertini}},
  \bibinfo{author}{\bibfnamefont{P.}~\bibnamefont{Kos}}, \bibnamefont{and}
  \bibinfo{author}{\bibfnamefont{T.~c.~v.} \bibnamefont{Prosen}},
  \bibinfo{journal}{Phys. Rev. X} \textbf{\bibinfo{volume}{9}},
  \bibinfo{pages}{021033} (\bibinfo{year}{2019}),
  \urlprefix\url{https://link.aps.org/doi/10.1103/PhysRevX.9.021033}.

\bibitem[{\citenamefont{{Parker} et~al.}(2018)\citenamefont{{Parker}, {Cao},
  {Avdoshkin}, {Scaffidi}, and {Altman}}}]{2018arXiv181208657P}
\bibinfo{author}{\bibfnamefont{D.~E.} \bibnamefont{{Parker}}},
  \bibinfo{author}{\bibfnamefont{X.}~\bibnamefont{{Cao}}},
  \bibinfo{author}{\bibfnamefont{A.}~\bibnamefont{{Avdoshkin}}},
  \bibinfo{author}{\bibfnamefont{T.}~\bibnamefont{{Scaffidi}}},
  \bibnamefont{and} \bibinfo{author}{\bibfnamefont{E.}~\bibnamefont{{Altman}}},
  \bibinfo{journal}{arXiv e-prints} \bibinfo{eid}{arXiv:1812.08657}
  (\bibinfo{year}{2018}), \eprint{1812.08657}.

\bibitem[{\citenamefont{Deutsch}(1991)}]{PhysRevA.43.2046}
\bibinfo{author}{\bibfnamefont{J.~M.} \bibnamefont{Deutsch}},
  \bibinfo{journal}{Phys. Rev. A} \textbf{\bibinfo{volume}{43}},
  \bibinfo{pages}{2046} (\bibinfo{year}{1991}),
  \urlprefix\url{http://link.aps.org/doi/10.1103/PhysRevA.43.2046}.

\bibitem[{\citenamefont{Srednicki}(1994)}]{PhysRevE.50.888}
\bibinfo{author}{\bibfnamefont{M.}~\bibnamefont{Srednicki}},
  \bibinfo{journal}{Phys. Rev. E} \textbf{\bibinfo{volume}{50}},
  \bibinfo{pages}{888} (\bibinfo{year}{1994}),
  \urlprefix\url{http://link.aps.org/doi/10.1103/PhysRevE.50.888}.

\bibitem[{\citenamefont{Luitz et~al.}(2015)\citenamefont{Luitz, Laflorencie,
  and Alet}}]{Luitz}
\bibinfo{author}{\bibfnamefont{D.~J.} \bibnamefont{Luitz}},
  \bibinfo{author}{\bibfnamefont{N.}~\bibnamefont{Laflorencie}},
  \bibnamefont{and} \bibinfo{author}{\bibfnamefont{F.}~\bibnamefont{Alet}},
  \bibinfo{journal}{Phys. Rev. B} \textbf{\bibinfo{volume}{91}},
  \bibinfo{pages}{081103} (\bibinfo{year}{2015}),
  \urlprefix\url{http://link.aps.org/doi/10.1103/PhysRevB.91.081103}.

\bibitem[{\citenamefont{Kj\"all et~al.}(2014)\citenamefont{Kj\"all, Bardarson,
  and Pollmann}}]{PhysRevLett.113.107204}
\bibinfo{author}{\bibfnamefont{J.~A.} \bibnamefont{Kj\"all}},
  \bibinfo{author}{\bibfnamefont{J.~H.} \bibnamefont{Bardarson}},
  \bibnamefont{and} \bibinfo{author}{\bibfnamefont{F.}~\bibnamefont{Pollmann}},
  \bibinfo{journal}{Phys. Rev. Lett.} \textbf{\bibinfo{volume}{113}},
  \bibinfo{pages}{107204} (\bibinfo{year}{2014}),
  \urlprefix\url{https://link.aps.org/doi/10.1103/PhysRevLett.113.107204}.

\bibitem[{\citenamefont{{Vosk} et~al.}(2014)\citenamefont{{Vosk}, {Huse}, and
  {Altman}}}]{VHA}
\bibinfo{author}{\bibfnamefont{R.}~\bibnamefont{{Vosk}}},
  \bibinfo{author}{\bibfnamefont{D.~A.} \bibnamefont{{Huse}}},
  \bibnamefont{and} \bibinfo{author}{\bibfnamefont{E.}~\bibnamefont{{Altman}}},
  \bibinfo{journal}{Phys. Rev. X} \textbf{\bibinfo{volume}{5}},
  \bibinfo{pages}{031032} (\bibinfo{year}{2014}), \eprint{1412.3117},
  \urlprefix\url{http://link.aps.org/doi/10.1103/PhysRevX.5.031032}.

\bibitem[{\citenamefont{{Potter} et~al.}(2015)\citenamefont{{Potter},
  {Vasseur}, and {Parameswaran}}}]{PVP}
\bibinfo{author}{\bibfnamefont{A.~C.} \bibnamefont{{Potter}}},
  \bibinfo{author}{\bibfnamefont{R.}~\bibnamefont{{Vasseur}}},
  \bibnamefont{and} \bibinfo{author}{\bibfnamefont{S.~A.}
  \bibnamefont{{Parameswaran}}}, \bibinfo{journal}{Phys. Rev. X}
  \textbf{\bibinfo{volume}{5}}, \bibinfo{pages}{031033} (\bibinfo{year}{2015}),
  \eprint{1501.03501},
  \urlprefix\url{http://link.aps.org/doi/10.1103/PhysRevX.5.031033}.

\bibitem[{\citenamefont{Serbyn et~al.}(2015)\citenamefont{Serbyn,
  Papi\ifmmode~\acute{c}\else \'{c}\fi{}, and Abanin}}]{PhysRevX.5.041047}
\bibinfo{author}{\bibfnamefont{M.}~\bibnamefont{Serbyn}},
  \bibinfo{author}{\bibfnamefont{Z.}~\bibnamefont{Papi\ifmmode~\acute{c}\else
  \'{c}\fi{}}}, \bibnamefont{and} \bibinfo{author}{\bibfnamefont{D.~A.}
  \bibnamefont{Abanin}}, \bibinfo{journal}{Physical Review X}
  \textbf{\bibinfo{volume}{5}}, \bibinfo{eid}{041047}
  (pages~\bibinfo{numpages}{10}) (\bibinfo{year}{2015}), \eprint{1507.01635},
  \urlprefix\url{http://link.aps.org/doi/10.1103/PhysRevX.5.041047}.

\bibitem[{\citenamefont{Khemani et~al.}(2017)\citenamefont{Khemani, Lim, Sheng,
  and Huse}}]{PhysRevX.7.021013}
\bibinfo{author}{\bibfnamefont{V.}~\bibnamefont{Khemani}},
  \bibinfo{author}{\bibfnamefont{S.~P.} \bibnamefont{Lim}},
  \bibinfo{author}{\bibfnamefont{D.~N.} \bibnamefont{Sheng}}, \bibnamefont{and}
  \bibinfo{author}{\bibfnamefont{D.~A.} \bibnamefont{Huse}},
  \bibinfo{journal}{Phys. Rev. X} \textbf{\bibinfo{volume}{7}},
  \bibinfo{pages}{021013} (\bibinfo{year}{2017}),
  \urlprefix\url{https://link.aps.org/doi/10.1103/PhysRevX.7.021013}.

\bibitem[{\citenamefont{Schreiber et~al.}(2015)\citenamefont{Schreiber,
  Hodgman, Bordia, L{\"u}schen, Fischer, Vosk, Altman, Schneider, and
  Bloch}}]{Schreiber842}
\bibinfo{author}{\bibfnamefont{M.}~\bibnamefont{Schreiber}},
  \bibinfo{author}{\bibfnamefont{S.~S.} \bibnamefont{Hodgman}},
  \bibinfo{author}{\bibfnamefont{P.}~\bibnamefont{Bordia}},
  \bibinfo{author}{\bibfnamefont{H.~P.} \bibnamefont{L{\"u}schen}},
  \bibinfo{author}{\bibfnamefont{M.~H.} \bibnamefont{Fischer}},
  \bibinfo{author}{\bibfnamefont{R.}~\bibnamefont{Vosk}},
  \bibinfo{author}{\bibfnamefont{E.}~\bibnamefont{Altman}},
  \bibinfo{author}{\bibfnamefont{U.}~\bibnamefont{Schneider}},
  \bibnamefont{and} \bibinfo{author}{\bibfnamefont{I.}~\bibnamefont{Bloch}},
  \bibinfo{journal}{Science} \textbf{\bibinfo{volume}{349}},
  \bibinfo{pages}{842} (\bibinfo{year}{2015}), ISSN \bibinfo{issn}{0036-8075},
  \eprint{https://science.sciencemag.org/content/349/6250/842.full.pdf},
  \urlprefix\url{https://science.sciencemag.org/content/349/6250/842}.

\bibitem[{\citenamefont{Dumitrescu et~al.}(2017)\citenamefont{Dumitrescu,
  Vasseur, and Potter}}]{DVP}
\bibinfo{author}{\bibfnamefont{P.~T.} \bibnamefont{Dumitrescu}},
  \bibinfo{author}{\bibfnamefont{R.}~\bibnamefont{Vasseur}}, \bibnamefont{and}
  \bibinfo{author}{\bibfnamefont{A.~C.} \bibnamefont{Potter}},
  \bibinfo{journal}{Phys. Rev. Lett.} \textbf{\bibinfo{volume}{119}},
  \bibinfo{pages}{110604} (\bibinfo{year}{2017}),
  \urlprefix\url{https://link.aps.org/doi/10.1103/PhysRevLett.119.110604}.

\bibitem[{\citenamefont{Thiery et~al.}(2018)\citenamefont{Thiery, Huveneers,
  M\"uller, and De~Roeck}}]{THMdR_meanfield}
\bibinfo{author}{\bibfnamefont{T.}~\bibnamefont{Thiery}},
  \bibinfo{author}{\bibfnamefont{F.}~\bibnamefont{Huveneers}},
  \bibinfo{author}{\bibfnamefont{M.}~\bibnamefont{M\"uller}}, \bibnamefont{and}
  \bibinfo{author}{\bibfnamefont{W.}~\bibnamefont{De~Roeck}},
  \bibinfo{journal}{Phys. Rev. Lett.} \textbf{\bibinfo{volume}{121}},
  \bibinfo{pages}{140601} (\bibinfo{year}{2018}),
  \urlprefix\url{https://link.aps.org/doi/10.1103/PhysRevLett.121.140601}.

\bibitem[{\citenamefont{Zhang et~al.}(2016)\citenamefont{Zhang, Zhao, Devakul,
  and Huse}}]{zhang_many-body_2016}
\bibinfo{author}{\bibfnamefont{L.}~\bibnamefont{Zhang}},
  \bibinfo{author}{\bibfnamefont{B.}~\bibnamefont{Zhao}},
  \bibinfo{author}{\bibfnamefont{T.}~\bibnamefont{Devakul}}, \bibnamefont{and}
  \bibinfo{author}{\bibfnamefont{D.~A.} \bibnamefont{Huse}},
  \bibinfo{journal}{Physical Review B} \textbf{\bibinfo{volume}{93}},
  \bibinfo{pages}{224201} (\bibinfo{year}{2016}),
  \urlprefix\url{https://link.aps.org/doi/10.1103/PhysRevB.93.224201}.

\bibitem[{\citenamefont{Goremykina et~al.}(2019)\citenamefont{Goremykina,
  Vasseur, and Serbyn}}]{goremykina_analytically_2019}
\bibinfo{author}{\bibfnamefont{A.}~\bibnamefont{Goremykina}},
  \bibinfo{author}{\bibfnamefont{R.}~\bibnamefont{Vasseur}}, \bibnamefont{and}
  \bibinfo{author}{\bibfnamefont{M.}~\bibnamefont{Serbyn}},
  \bibinfo{journal}{Physical Review Letters} \textbf{\bibinfo{volume}{122}},
  \bibinfo{pages}{040601} (\bibinfo{year}{2019}),
  \urlprefix\url{https://link.aps.org/doi/10.1103/PhysRevLett.122.040601}.

\bibitem[{\citenamefont{Dumitrescu et~al.}(2019)\citenamefont{Dumitrescu,
  Goremykina, Parameswaran, Serbyn, and Vasseur}}]{PhysRevB.99.094205}
\bibinfo{author}{\bibfnamefont{P.~T.} \bibnamefont{Dumitrescu}},
  \bibinfo{author}{\bibfnamefont{A.}~\bibnamefont{Goremykina}},
  \bibinfo{author}{\bibfnamefont{S.~A.} \bibnamefont{Parameswaran}},
  \bibinfo{author}{\bibfnamefont{M.}~\bibnamefont{Serbyn}}, \bibnamefont{and}
  \bibinfo{author}{\bibfnamefont{R.}~\bibnamefont{Vasseur}},
  \bibinfo{journal}{Phys. Rev. B} \textbf{\bibinfo{volume}{99}},
  \bibinfo{pages}{094205} (\bibinfo{year}{2019}),
  \urlprefix\url{https://link.aps.org/doi/10.1103/PhysRevB.99.094205}.

\bibitem[{\citenamefont{{Basko} et~al.}(2006)\citenamefont{{Basko}, {Aleiner},
  and {Altshuler}}}]{BAA}
\bibinfo{author}{\bibfnamefont{D.~M.} \bibnamefont{{Basko}}},
  \bibinfo{author}{\bibfnamefont{I.~L.} \bibnamefont{{Aleiner}}},
  \bibnamefont{and} \bibinfo{author}{\bibfnamefont{B.~L.}
  \bibnamefont{{Altshuler}}}, \bibinfo{journal}{Annals of Physics}
  \textbf{\bibinfo{volume}{321}}, \bibinfo{pages}{1126 }
  (\bibinfo{year}{2006}), ISSN \bibinfo{issn}{0003-4916},
  \eprint{cond-mat/0506617},
  \urlprefix\url{http://www.sciencedirect.com/science/article/pii/S0003491605002630}.

\bibitem[{\citenamefont{Oganesyan and Huse}(2007)}]{PhysRevB.75.155111}
\bibinfo{author}{\bibfnamefont{V.}~\bibnamefont{Oganesyan}} \bibnamefont{and}
  \bibinfo{author}{\bibfnamefont{D.~A.} \bibnamefont{Huse}},
  \bibinfo{journal}{Phys. Rev. B} \textbf{\bibinfo{volume}{75}},
  \bibinfo{eid}{155111} (pages~\bibinfo{numpages}{5}) (\bibinfo{year}{2007}),
  \eprint{cond-mat/0610854},
  \urlprefix\url{http://link.aps.org/doi/10.1103/PhysRevB.75.155111}.

\bibitem[{\citenamefont{Pal and Huse}(2010)}]{PalHuse}
\bibinfo{author}{\bibfnamefont{A.}~\bibnamefont{Pal}} \bibnamefont{and}
  \bibinfo{author}{\bibfnamefont{D.~A.} \bibnamefont{Huse}},
  \bibinfo{journal}{Phys. Rev. B} \textbf{\bibinfo{volume}{82}},
  \bibinfo{pages}{174411} (\bibinfo{year}{2010}),
  \urlprefix\url{http://link.aps.org/doi/10.1103/PhysRevB.82.174411}.

\bibitem[{\citenamefont{{Bauer} and {Nayak}}(2013)}]{BauerNayak}
\bibinfo{author}{\bibfnamefont{B.}~\bibnamefont{{Bauer}}} \bibnamefont{and}
  \bibinfo{author}{\bibfnamefont{C.}~\bibnamefont{{Nayak}}},
  \bibinfo{journal}{Journal of Statistical Mechanics: Theory and Experiment}
  \textbf{\bibinfo{volume}{2013}}, \bibinfo{eid}{09005} (\bibinfo{year}{2013}),
  \eprint{1306.5753},
  \urlprefix\url{http://stacks.iop.org/1742-5468/2013/i=09/a=P09005}.

\bibitem[{\citenamefont{Serbyn et~al.}(2013)\citenamefont{Serbyn,
  Papi\ifmmode~\acute{c}\else \'{c}\fi{}, and Abanin}}]{PhysRevLett.111.127201}
\bibinfo{author}{\bibfnamefont{M.}~\bibnamefont{Serbyn}},
  \bibinfo{author}{\bibfnamefont{Z.}~\bibnamefont{Papi\ifmmode~\acute{c}\else
  \'{c}\fi{}}}, \bibnamefont{and} \bibinfo{author}{\bibfnamefont{D.~A.}
  \bibnamefont{Abanin}}, \bibinfo{journal}{Physical Review Letters}
  \textbf{\bibinfo{volume}{111}}, \bibinfo{eid}{127201}
  (pages~\bibinfo{numpages}{5}) (\bibinfo{year}{2013}), \eprint{1305.5554},
  \urlprefix\url{link.aps.org/doi/10.1103/PhysRevLett.111.127201}.

\bibitem[{\citenamefont{{Huse} et~al.}(2014)\citenamefont{{Huse},
  {Nandkishore}, and {Oganesyan}}}]{PhysRevB.90.174202}
\bibinfo{author}{\bibfnamefont{D.~A.} \bibnamefont{{Huse}}},
  \bibinfo{author}{\bibfnamefont{R.}~\bibnamefont{{Nandkishore}}},
  \bibnamefont{and}
  \bibinfo{author}{\bibfnamefont{V.}~\bibnamefont{{Oganesyan}}},
  \bibinfo{journal}{Phys. Rev. B} \textbf{\bibinfo{volume}{90}},
  \bibinfo{eid}{174202} (pages~\bibinfo{numpages}{5}) (\bibinfo{year}{2014}),
  \eprint{1305.4915},
  \urlprefix\url{http://link.aps.org/doi/10.1103/PhysRevB.90.174202}.

\bibitem[{\citenamefont{Nandkishore and Huse}(2015)}]{2014arXiv1404.0686N}
\bibinfo{author}{\bibfnamefont{R.}~\bibnamefont{Nandkishore}} \bibnamefont{and}
  \bibinfo{author}{\bibfnamefont{D.~A.} \bibnamefont{Huse}},
  \bibinfo{journal}{Annual Review of Condensed Matter Physics}
  \textbf{\bibinfo{volume}{6}}, \bibinfo{pages}{15} (\bibinfo{year}{2015}),
  \eprint{1404.0686},
  \urlprefix\url{http://dx.doi.org/10.1146/annurev-conmatphys-031214-014726}.

\bibitem[{\citenamefont{Abanin et~al.}(2019)\citenamefont{Abanin, Altman,
  Bloch, and Serbyn}}]{RevModPhys.91.021001}
\bibinfo{author}{\bibfnamefont{D.~A.} \bibnamefont{Abanin}},
  \bibinfo{author}{\bibfnamefont{E.}~\bibnamefont{Altman}},
  \bibinfo{author}{\bibfnamefont{I.}~\bibnamefont{Bloch}}, \bibnamefont{and}
  \bibinfo{author}{\bibfnamefont{M.}~\bibnamefont{Serbyn}},
  \bibinfo{journal}{Rev. Mod. Phys.} \textbf{\bibinfo{volume}{91}},
  \bibinfo{pages}{021001} (\bibinfo{year}{2019}),
  \urlprefix\url{https://link.aps.org/doi/10.1103/RevModPhys.91.021001}.

\bibitem[{\citenamefont{Nahum et~al.}(2018)\citenamefont{Nahum, Vijay, and
  Haah}}]{PhysRevX.8.021014}
\bibinfo{author}{\bibfnamefont{A.}~\bibnamefont{Nahum}},
  \bibinfo{author}{\bibfnamefont{S.}~\bibnamefont{Vijay}}, \bibnamefont{and}
  \bibinfo{author}{\bibfnamefont{J.}~\bibnamefont{Haah}},
  \bibinfo{journal}{Phys. Rev. X} \textbf{\bibinfo{volume}{8}},
  \bibinfo{pages}{021014} (\bibinfo{year}{2018}),
  \urlprefix\url{https://link.aps.org/doi/10.1103/PhysRevX.8.021014}.

\bibitem[{\citenamefont{von Keyserlingk et~al.}(2018)\citenamefont{von
  Keyserlingk, Rakovszky, Pollmann, and Sondhi}}]{PhysRevX.8.021013}
\bibinfo{author}{\bibfnamefont{C.~W.} \bibnamefont{von Keyserlingk}},
  \bibinfo{author}{\bibfnamefont{T.}~\bibnamefont{Rakovszky}},
  \bibinfo{author}{\bibfnamefont{F.}~\bibnamefont{Pollmann}}, \bibnamefont{and}
  \bibinfo{author}{\bibfnamefont{S.~L.} \bibnamefont{Sondhi}},
  \bibinfo{journal}{Phys. Rev. X} \textbf{\bibinfo{volume}{8}},
  \bibinfo{pages}{021013} (\bibinfo{year}{2018}),
  \urlprefix\url{https://link.aps.org/doi/10.1103/PhysRevX.8.021013}.

\bibitem[{\citenamefont{Chan et~al.}(2018{\natexlab{a}})\citenamefont{Chan,
  De~Luca, and Chalker}}]{PhysRevX.8.041019}
\bibinfo{author}{\bibfnamefont{A.}~\bibnamefont{Chan}},
  \bibinfo{author}{\bibfnamefont{A.}~\bibnamefont{De~Luca}}, \bibnamefont{and}
  \bibinfo{author}{\bibfnamefont{J.~T.} \bibnamefont{Chalker}},
  \bibinfo{journal}{Phys. Rev. X} \textbf{\bibinfo{volume}{8}},
  \bibinfo{pages}{041019} (\bibinfo{year}{2018}{\natexlab{a}}),
  \urlprefix\url{https://link.aps.org/doi/10.1103/PhysRevX.8.041019}.

\bibitem[{\citenamefont{Chan et~al.}(2018{\natexlab{b}})\citenamefont{Chan,
  De~Luca, and Chalker}}]{PhysRevLett.121.060601}
\bibinfo{author}{\bibfnamefont{A.}~\bibnamefont{Chan}},
  \bibinfo{author}{\bibfnamefont{A.}~\bibnamefont{De~Luca}}, \bibnamefont{and}
  \bibinfo{author}{\bibfnamefont{J.~T.} \bibnamefont{Chalker}},
  \bibinfo{journal}{Phys. Rev. Lett.} \textbf{\bibinfo{volume}{121}},
  \bibinfo{pages}{060601} (\bibinfo{year}{2018}{\natexlab{b}}),
  \urlprefix\url{https://link.aps.org/doi/10.1103/PhysRevLett.121.060601}.

\bibitem[{\citenamefont{Rakovszky et~al.}(2018)\citenamefont{Rakovszky,
  Pollmann, and von Keyserlingk}}]{PhysRevX.8.031058}
\bibinfo{author}{\bibfnamefont{T.}~\bibnamefont{Rakovszky}},
  \bibinfo{author}{\bibfnamefont{F.}~\bibnamefont{Pollmann}}, \bibnamefont{and}
  \bibinfo{author}{\bibfnamefont{C.~W.} \bibnamefont{von Keyserlingk}},
  \bibinfo{journal}{Phys. Rev. X} \textbf{\bibinfo{volume}{8}},
  \bibinfo{pages}{031058} (\bibinfo{year}{2018}),
  \urlprefix\url{https://link.aps.org/doi/10.1103/PhysRevX.8.031058}.

\bibitem[{\citenamefont{Zhou and Nahum}(2019)}]{PhysRevB.99.174205}
\bibinfo{author}{\bibfnamefont{T.}~\bibnamefont{Zhou}} \bibnamefont{and}
  \bibinfo{author}{\bibfnamefont{A.}~\bibnamefont{Nahum}},
  \bibinfo{journal}{Phys. Rev. B} \textbf{\bibinfo{volume}{99}},
  \bibinfo{pages}{174205} (\bibinfo{year}{2019}),
  \urlprefix\url{https://link.aps.org/doi/10.1103/PhysRevB.99.174205}.

\bibitem[{\citenamefont{Khemani et~al.}(2018)\citenamefont{Khemani, Vishwanath,
  and Huse}}]{PhysRevX.8.031057}
\bibinfo{author}{\bibfnamefont{V.}~\bibnamefont{Khemani}},
  \bibinfo{author}{\bibfnamefont{A.}~\bibnamefont{Vishwanath}},
  \bibnamefont{and} \bibinfo{author}{\bibfnamefont{D.~A.} \bibnamefont{Huse}},
  \bibinfo{journal}{Phys. Rev. X} \textbf{\bibinfo{volume}{8}},
  \bibinfo{pages}{031057} (\bibinfo{year}{2018}),
  \urlprefix\url{https://link.aps.org/doi/10.1103/PhysRevX.8.031057}.

\bibitem[{\citenamefont{{Friedman} et~al.}(2019)\citenamefont{{Friedman},
  {Chan}, {De Luca}, and {Chalker}}}]{2019arXiv190607736F}
\bibinfo{author}{\bibfnamefont{A.~J.} \bibnamefont{{Friedman}}},
  \bibinfo{author}{\bibfnamefont{A.}~\bibnamefont{{Chan}}},
  \bibinfo{author}{\bibfnamefont{A.}~\bibnamefont{{De Luca}}},
  \bibnamefont{and} \bibinfo{author}{\bibfnamefont{J.~T.}
  \bibnamefont{{Chalker}}}, \bibinfo{journal}{arXiv e-prints}
  \bibinfo{eid}{arXiv:1906.07736} (\bibinfo{year}{2019}), \eprint{1906.07736}.

\bibitem[{\citenamefont{{Skinner} et~al.}(2018)\citenamefont{{Skinner},
  {Ruhman}, and {Nahum}}}]{Skinner2018}
\bibinfo{author}{\bibfnamefont{B.}~\bibnamefont{{Skinner}}},
  \bibinfo{author}{\bibfnamefont{J.}~\bibnamefont{{Ruhman}}}, \bibnamefont{and}
  \bibinfo{author}{\bibfnamefont{A.}~\bibnamefont{{Nahum}}},
  \bibinfo{journal}{arXiv e-prints} \bibinfo{eid}{arXiv:1808.05953}
  (\bibinfo{year}{2018}), \eprint{1808.05953}.

\bibitem[{\citenamefont{{Li} et~al.}(2018)\citenamefont{{Li}, {Chen}, and
  {Fisher}}}]{LiChenFisher2018}
\bibinfo{author}{\bibfnamefont{Y.}~\bibnamefont{{Li}}},
  \bibinfo{author}{\bibfnamefont{X.}~\bibnamefont{{Chen}}}, \bibnamefont{and}
  \bibinfo{author}{\bibfnamefont{M.~P.~A.} \bibnamefont{{Fisher}}},
  \bibinfo{journal}{\prb} \textbf{\bibinfo{volume}{98}}, \bibinfo{eid}{205136}
  (\bibinfo{year}{2018}), \eprint{1808.06134}.

\bibitem[{\citenamefont{{Chan} et~al.}(2019)\citenamefont{{Chan},
  {Nandkishore}, {Pretko}, and {Smith}}}]{Chan2019}
\bibinfo{author}{\bibfnamefont{A.}~\bibnamefont{{Chan}}},
  \bibinfo{author}{\bibfnamefont{R.~M.} \bibnamefont{{Nandkishore}}},
  \bibinfo{author}{\bibfnamefont{M.}~\bibnamefont{{Pretko}}}, \bibnamefont{and}
  \bibinfo{author}{\bibfnamefont{G.}~\bibnamefont{{Smith}}},
  \bibinfo{journal}{\prb} \textbf{\bibinfo{volume}{99}}, \bibinfo{eid}{224307}
  (\bibinfo{year}{2019}), \eprint{1808.05949}.

\bibitem[{\citenamefont{{Li} et~al.}(2019)\citenamefont{{Li}, {Chen}, and
  {Fisher}}}]{LiChenFisher2019}
\bibinfo{author}{\bibfnamefont{Y.}~\bibnamefont{{Li}}},
  \bibinfo{author}{\bibfnamefont{X.}~\bibnamefont{{Chen}}}, \bibnamefont{and}
  \bibinfo{author}{\bibfnamefont{M.~P.~A.} \bibnamefont{{Fisher}}},
  \bibinfo{journal}{arXiv e-prints} \bibinfo{eid}{arXiv:1901.08092}
  (\bibinfo{year}{2019}), \eprint{1901.08092}.

\bibitem[{\citenamefont{{Choi} et~al.}(2019)\citenamefont{{Choi}, {Bao}, {Qi},
  and {Altman}}}]{2019arXiv190305124C}
\bibinfo{author}{\bibfnamefont{S.}~\bibnamefont{{Choi}}},
  \bibinfo{author}{\bibfnamefont{Y.}~\bibnamefont{{Bao}}},
  \bibinfo{author}{\bibfnamefont{X.-L.} \bibnamefont{{Qi}}}, \bibnamefont{and}
  \bibinfo{author}{\bibfnamefont{E.}~\bibnamefont{{Altman}}},
  \bibinfo{journal}{arXiv e-prints} \bibinfo{eid}{arXiv:1903.05124}
  (\bibinfo{year}{2019}), \eprint{1903.05124}.

\bibitem[{\citenamefont{{Szyniszewski}
  et~al.}(2019)\citenamefont{{Szyniszewski}, {Romito}, and
  {Schomerus}}}]{2019arXiv190305452S}
\bibinfo{author}{\bibfnamefont{M.}~\bibnamefont{{Szyniszewski}}},
  \bibinfo{author}{\bibfnamefont{A.}~\bibnamefont{{Romito}}}, \bibnamefont{and}
  \bibinfo{author}{\bibfnamefont{H.}~\bibnamefont{{Schomerus}}},
  \bibinfo{journal}{arXiv e-prints} \bibinfo{eid}{arXiv:1903.05452}
  (\bibinfo{year}{2019}), \eprint{1903.05452}.

\bibitem[{\citenamefont{{Gullans} and {Huse}}(2019)}]{2019arXiv190505195G}
\bibinfo{author}{\bibfnamefont{M.~J.} \bibnamefont{{Gullans}}}
  \bibnamefont{and} \bibinfo{author}{\bibfnamefont{D.~A.}
  \bibnamefont{{Huse}}}, \bibinfo{journal}{arXiv e-prints}
  \bibinfo{eid}{arXiv:1905.05195} (\bibinfo{year}{2019}), \eprint{1905.05195}.

\bibitem[{\citenamefont{{Hayden} et~al.}(2016)\citenamefont{{Hayden}, {Nezami},
  {Qi}, {Thomas}, {Walter}, and {Yang}}}]{Hayden2016}
\bibinfo{author}{\bibfnamefont{P.}~\bibnamefont{{Hayden}}},
  \bibinfo{author}{\bibfnamefont{S.}~\bibnamefont{{Nezami}}},
  \bibinfo{author}{\bibfnamefont{X.-L.} \bibnamefont{{Qi}}},
  \bibinfo{author}{\bibfnamefont{N.}~\bibnamefont{{Thomas}}},
  \bibinfo{author}{\bibfnamefont{M.}~\bibnamefont{{Walter}}}, \bibnamefont{and}
  \bibinfo{author}{\bibfnamefont{Z.}~\bibnamefont{{Yang}}},
  \bibinfo{journal}{JHEP} \textbf{\bibinfo{volume}{2016}}, \bibinfo{pages}{9}
  (\bibinfo{year}{2016}), ISSN \bibinfo{issn}{1029-8479}, \eprint{1601.01694},
  \urlprefix\url{http://dx.doi.org/10.1007/JHEP11(2016)009}.

\bibitem[{\citenamefont{Qi et~al.}(2017)\citenamefont{Qi, Yang, and
  You}}]{Qi:2017qf}
\bibinfo{author}{\bibfnamefont{X.-L.} \bibnamefont{Qi}},
  \bibinfo{author}{\bibfnamefont{Z.}~\bibnamefont{Yang}}, \bibnamefont{and}
  \bibinfo{author}{\bibfnamefont{Y.-Z.} \bibnamefont{You}},
  \bibinfo{journal}{Journal of High Energy Physics}
  \textbf{\bibinfo{volume}{2017}}, \bibinfo{pages}{60} (\bibinfo{year}{2017}),
  ISSN \bibinfo{issn}{1029-8479},
  \urlprefix\url{https://doi.org/10.1007/JHEP08(2017)060}.

\bibitem[{\citenamefont{Nielsen and Chuang}(2010)}]{NielsenChuang}
\bibinfo{author}{\bibfnamefont{M.~A.} \bibnamefont{Nielsen}} \bibnamefont{and}
  \bibinfo{author}{\bibfnamefont{I.~L.} \bibnamefont{Chuang}},
  \emph{\bibinfo{title}{Computation and Quantum Information}}
  (\bibinfo{publisher}{Cambridge University Press},
  \bibinfo{address}{Cambridge, UK}, \bibinfo{year}{2010}).

\bibitem[{\citenamefont{{Wiseman}}(1996)}]{Wiseman1996}
\bibinfo{author}{\bibfnamefont{H.~M.} \bibnamefont{{Wiseman}}},
  \bibinfo{journal}{Quantum and Semiclassical Optics}
  \textbf{\bibinfo{volume}{8}}, \bibinfo{pages}{205} (\bibinfo{year}{1996}),
  \eprint{quant-ph/0302080}.

\bibitem[{\citenamefont{Or{\'u}s}(2014)}]{ORUS2014117}
\bibinfo{author}{\bibfnamefont{R.}~\bibnamefont{Or{\'u}s}},
  \bibinfo{journal}{Annals of Physics} \textbf{\bibinfo{volume}{349}},
  \bibinfo{pages}{117 } (\bibinfo{year}{2014}), ISSN \bibinfo{issn}{0003-4916},
  \urlprefix\url{http://www.sciencedirect.com/science/article/pii/S0003491614001596}.

\bibitem[{\citenamefont{Cardy}(1984)}]{cardy_conformal_1984}
\bibinfo{author}{\bibfnamefont{J.~L.} \bibnamefont{Cardy}},
  \bibinfo{journal}{Nuclear Physics B} \textbf{\bibinfo{volume}{240}},
  \bibinfo{pages}{514} (\bibinfo{year}{1984}), ISSN \bibinfo{issn}{0550-3213},
  \urlprefix\url{http://www.sciencedirect.com/science/article/pii/0550321384902414}.

\bibitem[{\citenamefont{Cardy}(2006)}]{cardy_boundary_2006}
\bibinfo{author}{\bibfnamefont{J.}~\bibnamefont{Cardy}},
  \bibinfo{journal}{Encyclopedia of Mathematical Physics}
  (\bibinfo{year}{2006}), \bibinfo{note}{arXiv: hep-th/0411189},
  \urlprefix\url{http://arxiv.org/abs/hep-th/0411189}.

\bibitem[{\citenamefont{Francesco et~al.}(2012)\citenamefont{Francesco,
  Mathieu, and S{\'e}n{\'e}chal}}]{francesco2012conformal}
\bibinfo{author}{\bibfnamefont{P.}~\bibnamefont{Francesco}},
  \bibinfo{author}{\bibfnamefont{P.}~\bibnamefont{Mathieu}}, \bibnamefont{and}
  \bibinfo{author}{\bibfnamefont{D.}~\bibnamefont{S{\'e}n{\'e}chal}},
  \emph{\bibinfo{title}{Conformal field theory}} (\bibinfo{publisher}{Springer
  Science \& Business Media}, \bibinfo{year}{2012}).

\bibitem[{\citenamefont{Fortuin and Kasteleyn}(1972)}]{FORTUIN1972536}
\bibinfo{author}{\bibfnamefont{C.}~\bibnamefont{Fortuin}} \bibnamefont{and}
  \bibinfo{author}{\bibfnamefont{P.}~\bibnamefont{Kasteleyn}},
  \bibinfo{journal}{Physica} \textbf{\bibinfo{volume}{57}}, \bibinfo{pages}{536
  } (\bibinfo{year}{1972}), ISSN \bibinfo{issn}{0031-8914},
  \urlprefix\url{http://www.sciencedirect.com/science/article/pii/0031891472900456}.

\bibitem[{\citenamefont{Black and Emery}(1981)}]{PhysRevB.23.429}
\bibinfo{author}{\bibfnamefont{J.~L.} \bibnamefont{Black}} \bibnamefont{and}
  \bibinfo{author}{\bibfnamefont{V.~J.} \bibnamefont{Emery}},
  \bibinfo{journal}{Phys. Rev. B} \textbf{\bibinfo{volume}{23}},
  \bibinfo{pages}{429} (\bibinfo{year}{1981}),
  \urlprefix\url{https://link.aps.org/doi/10.1103/PhysRevB.23.429}.

\bibitem[{\citenamefont{Dubail et~al.}(2010)\citenamefont{Dubail, Jacobsen, and
  Saleur}}]{Dubail_2010}
\bibinfo{author}{\bibfnamefont{J.}~\bibnamefont{Dubail}},
  \bibinfo{author}{\bibfnamefont{J.~L.} \bibnamefont{Jacobsen}},
  \bibnamefont{and} \bibinfo{author}{\bibfnamefont{H.}~\bibnamefont{Saleur}},
  \bibinfo{journal}{Journal of Statistical Mechanics: Theory and Experiment}
  \textbf{\bibinfo{volume}{2010}}, \bibinfo{pages}{P12026}
  (\bibinfo{year}{2010}),
  \urlprefix\url{https://doi.org/10.1088%2F1742-5468%2F2010%2F12%2Fp12026}.

\bibitem[{\citenamefont{Cardy}(1989)}]{CARDY1989581}
\bibinfo{author}{\bibfnamefont{J.~L.} \bibnamefont{Cardy}},
  \bibinfo{journal}{Nuclear Physics B} \textbf{\bibinfo{volume}{324}},
  \bibinfo{pages}{581 } (\bibinfo{year}{1989}), ISSN \bibinfo{issn}{0550-3213},
  \urlprefix\url{http://www.sciencedirect.com/science/article/pii/055032138990521X}.

\bibitem[{\citenamefont{Cardy}(1992)}]{Cardy_1992}
\bibinfo{author}{\bibfnamefont{J.~L.} \bibnamefont{Cardy}},
  \bibinfo{journal}{Journal of Physics A: Mathematical and General}
  \textbf{\bibinfo{volume}{25}}, \bibinfo{pages}{L201} (\bibinfo{year}{1992}),
  \urlprefix\url{https://doi.org/10.1088%2F0305-4470%2F25%2F4%2F009}.

\bibitem[{\citenamefont{den Nijs}(1983)}]{PhysRevB.27.1674}
\bibinfo{author}{\bibfnamefont{M.}~\bibnamefont{den Nijs}},
  \bibinfo{journal}{Phys. Rev. B} \textbf{\bibinfo{volume}{27}},
  \bibinfo{pages}{1674} (\bibinfo{year}{1983}),
  \urlprefix\url{https://link.aps.org/doi/10.1103/PhysRevB.27.1674}.

\bibitem[{\citenamefont{{Bao} et~al.}(2019)\citenamefont{{Bao}, {Choi}, and
  {Altman}}}]{2019arXiv190804305B}
\bibinfo{author}{\bibfnamefont{Y.}~\bibnamefont{{Bao}}},
  \bibinfo{author}{\bibfnamefont{S.}~\bibnamefont{{Choi}}}, \bibnamefont{and}
  \bibinfo{author}{\bibfnamefont{E.}~\bibnamefont{{Altman}}},
  \bibinfo{journal}{arXiv e-prints} \bibinfo{eid}{arXiv:1908.04305}
  (\bibinfo{year}{2019}), \eprint{1908.04305}.

\end{thebibliography}

\newpage
\onecolumngrid
\appendix
\section{The $1/d$ Expansion}\label{app:1/d}

In this appendix, we derive the $1/d$ expansion of the triangle weight in \eqnref{eq:J def}. Let $|g|$ be the number of cycles in the permutation $g$ for $g\in S_Q$. We define the following group functions: $\delta_g$ and $\delta'_g$,
\begin{equation}
\delta_{g}=\left\{\begin{array}{ll}
1& \text{if $|g|=Q$, i.e.~$g=e$}, \\
0& \text{otherwise};
\end{array}\right.
\qquad
\delta'_{g}=\left\{\begin{array}{ll}
1& \text{if $|g|=Q-1$, i.e.~$g$ is a transposition}, \\
0& \text{otherwise}.
\end{array}\right.
\end{equation}
Then the function $d^{|g|}$ can be expanded as
\begin{equation}\label{eq:d^g to 1/d}
d^{|g|}=d^{Q}\delta_{g}+d^{Q-1}\delta'_{g}+\cdots
=d^{Q}\left(\delta_{g}+d^{-1}\delta'_{g}+\scO(d^{-2})\right).
\end{equation}
With \eqnref{eq:d^g to 1/d}, the weight function $W_p(g)$ in \eqnref{LabelEqGenericBoltzmannWeight} admits the following expansion,
\begin{equation}\label{eq:Wp to 1/d}
W_p(g)=(1-p)d^{|g|}+p d^Q
=d^{Q}\left((1-p)\delta_{g}+p+\frac{1-p}{d}\delta'_{g}+\scO(d^{-2})\right).
\end{equation}

The Weingarten function $\mathsf{Wg}_D(g)$ in \eqnref{eq:Wg def} (with $D=d^2$) has an alternative definition that it is a class function satisfying the following equation
\begin{equation}\label{eq:Wg equation}
\sum_{g_2\in S_Q}\mathsf{Wg}_D(g_1^{-1}g_2) D^{|g_2^{-1}g_3|}=\delta_{g_1^{-1}g_3}.
\end{equation}
Given \eqnref{eq:d^g to 1/d}, one can verify that the following expansion
\begin{equation}\label{eq:Wg to 1/d}
\mathsf{Wg}_D(g)=D^{-Q}\left(\delta_{g}-D^{-1}\delta'_{g}+\scO(D^{-2})\right)
\end{equation}
is a solution of \eqnref{eq:Wg equation} to the order of $1/D$, as
\begin{equation}
\begin{split}
\sum_{g_2\in S_Q}\mathsf{Wg}_D(g_1^{-1}g_2) D^{|g_2^{-1}g_3|}&=\sum_{g_2\in S_Q}\left(\delta_{g_1^{-1}g_2}-D^{-1}\delta'_{g_1^{-1}g_2}+\scO(D^{-2})\right) \left(\delta_{g_2^{-1}g_3}+D^{-1}\delta'_{g_2^{-1}g_3}+\scO(D^{-2})\right)\\
&=\delta_{g_1^{-1}g_3}+\scO(D^{-2}).
\end{split}
\end{equation}

Using \eqnref{eq:Wp to 1/d} and \eqnref{eq:Wg to 1/d}, we can now evaluate the $1/d$ expansion for the triangle weight $J(g_i,g_j;g_k)$,
\begin{equation}
\begin{split}
J_p(g_i,g_j;g_k)&=\sum_{g_l\in S_Q}W_p(g_i^{-1}g_l)W_p(g_j^{-1}g_l)\mathsf{Wg}_{d^2}(g_l^{-1}g_k)\\
&=\sum_{g_l\in S_Q}d^{Q}\left((1-p)\delta_{g_i^{-1}g_l}+p+\frac{1-p}{d}\delta'_{g_i^{-1}g_l}+\scO(d^{-2})\right)\\
&\phantom{=\sum_{g_l\in S_Q}}d^{Q}\left((1-p)\delta_{g_j^{-1}g_l}+p+\frac{1-p}{d}\delta'_{g_j^{-1}g_l}+\scO(d^{-2})\right)\\
&\phantom{=\sum_{g_l\in S_Q}}d^{-2Q}\left(\delta_{g_l^{-1}g_k}-d^{-2}\delta'_{g_l^{-1}g_k}+\scO(d^{-4})\right)\\
&=\left((1-p)\delta_{g_i^{-1}g_k}+p\right)\left((1-p)\delta_{g_j^{-1}g_k}+p\right)\\
&\phantom{=}+\left((1-p)\delta_{g_i^{-1}g_k}+p\right)\frac{1-p}{d}\delta'_{g_j^{-1}g_k}+\left((1-p)\delta_{g_j^{-1}g_k}+p\right)\frac{1-p}{d}\delta'_{g_i^{-1}g_k}+\scO(d^{-2})\\
&=\left((1-p)\delta_{g_i^{-1}g_k}+p\right)\left((1-p)\delta_{g_j^{-1}g_k}+p\right)\\
&\phantom{=}+\frac{1-p}{d}\left((1-p)\delta'_{g_i^{-1}g_j}(\delta_{g_i^{-1}g_k}
+\delta_{g_j^{-1}g_k})
+p(\delta'_{g_j^{-1}g_k}+\delta'_{g_i^{-1}g_k})\right)+\scO(d^{-2}).
\end{split}
\end{equation}
The result matches \eqnref{eq:J inf d} and \eqnref{eq:J to 1/d} as claimed in the main text.

To the same order $ \scO(d^{-2})$, this triangle weight can be rewritten as a product of Boltzmann weight on the links of the square lattice in Fig.~\ref{fig:lattice} (b):
\begin{equation}
J_p(g_i,g_j;g_k) = \left((1-p)\delta_{g_i^{-1}g_k}+p + \frac{1-p}{d}\delta'_{g_i^{-1}g_k} \right)\left((1-p)\delta_{g_j^{-1}g_k}+p + \frac{1-p}{d}\delta'_{g_j^{-1}g_k} \right) + \scO(d^{-2}).
\end{equation}
This can be checked by expanding the product explicitly and using the identity $\delta_{g_i^{-1}g_k} \delta'_{g_j^{-1}g_k} = \delta_{g_i^{-1}g_k} \delta'_{g_j^{-1}g_i}$. Therefore, at this order, the weights of the $S_Q$ model can be factorized into a product of local weights over links of the square lattice, that we rewrite as $e^{- E(g_i, g_k)}$, with the energy function
\begin{equation}
E(g_i,g_k) = - \log \left[ p + (1-p) \left(\delta_{g_i^{-1}g_k}+ \frac{1}{d}\delta'_{g_i^{-1}g_k} \right)
+ \scO(d^{-2})
\right].
\end{equation}

\end{document}